\documentclass[utf8, a4, 12pt]{article}

\RequirePackage{natbib}
\usepackage[top=30pt,bottom=30pt,left=40pt,right=48pt, marginpar=0pt, bindingoffset=0pt]{geometry} %

\RequirePackage[hyphens]{url}
\usepackage{url,hyperref,lineno,microtype,subcaption}
\hypersetup{breaklinks=true}
\usepackage{hhline}
\usepackage[onehalfspacing]{setspace}
\usepackage{titlesec} %

\usepackage[usenames, dvipsnames, table]{xcolor} %
\usepackage{amsfonts} %
\RequirePackage{inputenc,crop,graphicx,amsmath,array,color,amssymb,flushend,stfloats,amsthm,chngpage,times,datetime,parskip,epstopdf}

\usepackage{multirow}
\usepackage{booktabs}
\usepackage{enumerate}

\definecolor{yescolor}{RGB}{0,204,51}
\definecolor{nocolor}{RGB}{255, 77, 77}
\definecolor{maybecolor}{RGB}{255, 235, 153}
\definecolor{maybecolor}{RGB}{255, 235, 153}
\definecolor{partcolor}{RGB}{255, 133, 51}

\newcommand{\yes}{\color{PineGreen}$\checkmark$}
\newcommand{\no}{\color{red}$\times$}

\newcommand{\inpart}{\color{partcolor}}

\newcommand{\PreserveBackslash}[1]{\let\temp=\\#1\let\\=\temp}
\newcolumntype{C}[1]{>{\PreserveBackslash\centering}m{#1}}
\newcolumntype{R}[1]{>{\PreserveBackslash\raggedleft}m{#1}}
\newcolumntype{L}[1]{>{\PreserveBackslash\raggedright}m{#1}}

\makeatletter
\renewenvironment{abstract}{%
    \if@twocolumn
      \section*{\abstractname}%
    \else %
      \begin{center}%
        {\bfseries \Large\abstractname\vspace{\z@}}%
      \end{center}%
      \quotation
    \fi}
    {\if@twocolumn\else\endquotation\fi}
\makeatother

\titleformat{\paragraph}[hang]{\normalfont\normalsize\bfseries}{\theparagraph}{1em}{}
\titlespacing*{\paragraph} {0pt}{3.25ex plus 1ex minus .2ex}{1em}

\newcommand{\insilico}{{\emph{in silico~}}}

\def\Authors{ 
\hspace*{-\parindent}
Sotirios Panagiotou\,$^{1,2}$, Harry Sidiropoulos\,$^{2}$, \\ Mario Negrello\,$^{2}$, Dimitrios Soudris\,$^{1}$ and Christos Strydis\,$^{2,3}$}
\def\Address{
$^{1}$Microprocessors and Digital Systems Laboratory, National Technical University of Athens, Athens, Greece \\
$^{2}$Neurocomputing Laboratory, Erasmus Medical Centre, Department of Neuroscience, Rotterdam, The Netherlands \\
$^{3}$ Quantum and Computer Engineering Department, Delft University of Technology, Delft, The Netherlands
}

\begin{document}

\onecolumn
\title{EDEN: A high-performance, general-purpose, NeuroML-based neural simulator}

\author{\Authors} %

\date{}
\maketitle

{

\begin{quote}
    \Address{} %
\end{quote}

}

\begin{abstract}

Modern neuroscience employs in silico experimentation on ever-increasing and more detailed neural networks. The high modelling detail goes hand in hand with the need for high model reproducibility, reusability and transparency. Besides, the size of the models and the long timescales under study mandate the use of a simulation system with high computational performance, so as to provide an acceptable time to result. In this work, we present EDEN (Extensible Dynamics Engine for Networks), a new general-purpose, NeuroML-based neural simulator that achieves both high model flexibility and high computational performance, through an innovative model-analysis and code-generation technique.
The simulator 
runs NeuroML v2 models directly, eliminating the need for users to learn yet another simulator-specific, model-specification language. EDEN's functional correctness and computational performance were assessed through NeuroML models available on the NeuroML-DB and Open Source Brain model repositories. In qualitative experiments, the results produced by EDEN were verified against the established NEURON simulator, for a wide range of models. At the same time, computational-performance benchmarks reveal that EDEN runs up to 2 orders-of-magnitude faster than NEURON on a typical desktop computer, and does so without additional effort from the user. Finally, and without added user effort, EDEN has been built from scratch to scale seamlessly over multiple CPUs and across computer clusters, when available.

\tiny
\end{abstract}

\section{Introduction} \label{sec:intro}

Simulation of biological neural networks is an essential tool of modern neuroscience. However, there are currently certain challenges associated with the development and \insilico study of such networks. The neural models in use are diverse and heterogeneous; there is no single set of mathematical formulae that is commonly used by the majority of models. In addition, the biophysical mechanisms that make up models are constantly being modified, and reused in various combinations in new models. These factors mandate the use of general-purpose neural simulators in common practice. At the same time, the network sizes and levels of model detail employed by modern neuroscience translate to a constant increase in the amount of required computations. Thus, neuroscience projects need high-performance tools for simulations to finish in a practical time frame and for models to fit in available computer memory.

Although there already exists a rich arsenal of simulators targeting neuroscience, the aforementioned challenges of neural simulation remain an open problem. On one hand, there are hand-written codes that push the processing hardware to the limit but they are difficult or impossible to extend in terms of model support, because of their over-specialisation. They offer great computational performance by executing solely the numerical calculations required by the model's dynamics. On the other hand, there are general-purpose simulators that readily support most types of models, however, their computational efficiency is much smaller than that of hand-written codes. Hence, there is a significant gap in efficiency between general-purpose neural simulators and the computational capabilities that modern hardware platforms can achieve.

Besides, simulation of large networks often requires deploying neural models on multiple processor cores or, even, on computer clusters. Existing general-purpose simulators do not manage the technicalities of parallelisation, model decomposition and communication automatically. Thus, significant engineering effort is spent on setting up the simulators to run on multi-core and multi-node systems, which further obstructs scientific work.

A final problem is that, presently, each neural simulator uses its own model-specification language. Thus, models written for one simulator are difficult and laborious to adapt for another, which hampers the exchange and reuse of models across the neuroscience community. In this context, if a new simulator were to support only its own modelling language, this would fragment the modelling community further and would add a serious barrier to the simulator's adoption as well as the reuse of existing models.

\subsection{The EDEN simulator}

To address the challenges in \insilico neuroscience, we designed a new general-purpose neural simulator, called EDEN (Extensible Dynamics Engine for Networks). EDEN directly runs models described in NeuroML, achieves leading computational performance through a novel architecture, and handles parallel-processing resources -- both on standalone personal computers as well as on computer clusters -- automatically.

EDEN employs an innovative \emph{model-analysis} and \emph{code-generation technique} through which the model's variables and the mathematical operations needed for the simulation are converted into a set of individual \emph{work items}. Each work item consists of the data that represent a part of the neural network, and the calculations to simulate this part of the network over time. The calculations for the individual work items can then be run \emph{in parallel} within each simulation step, allowing distribution of the computational load among many processing elements. This technique enables \emph{by-design support} for general neural models, and at the same time offers significant performance benefits over conventional approaches.
The need for model generality with user-provided formulae is \emph{directly} addressed via \emph{automatic code generation}; but the architecture also supports hand-optimised implementations that apply for specific types of neurons.
At the same time, reducing the complex structure of biophysical mechanisms inside a neuron into an explicitly laid out set of 
essential, model-specific calculations allows compilers to perform large-scale optimisations. What is more, traditional simulators perform best with specific kinds of neuron models (e.g. multi-compartmental or point neurons) and worse with others. In contrast, EDEN's approach allows selecting the implementation that works best for each part of the network, at run time.

We adopted the \emph{NeuroML v2} standard~\citep{lems} as our simulator's modelling language. NeuroML v2 is the emerging standard cross-tool specification language for general neural-network models. By following the standard, we stay compatible with the entire NeuroML-software ecosystem: EDEN's simulation functionality is complemented by all the existing model-generation and results-analysis tools, and the ecosystem gets the most value out of EDEN as an interoperable simulator. Furthermore, positioning the simulator as a plug-compatible tool in the NeuroML stack allows us to focus our efforts on EDEN's features as a simulator (namely, computational performance, model generality and usability). Finally, supporting an established modelling language makes user adoption much easier, compared to introducing a new simulator-specific language.

Another aspect that was taken into account in EDEN's design is \emph{usability}. In addition to the benefits gained through NeuroML support, EDEN addresses usability through \emph{automatic management} of multi-processing resources. This means that EDEN can distribute processing for a simulation across the processor cores of one personal computer -- or even a computer cluster -- fully automatically. Thus, users can fully exploit their modern computer hardware and deploy simulations of large networks on high-performance clusters, with no additional effort.

To evaluate all aforementioned features of EDEN, we employed: (1) qualitative benchmarks showing simulation fidelity to the standard neural simulator NEURON; and (2) quantitative benchmarks showing far superior simulation speed compared to NEURON, for networks of non-trivial size. The results of these benchmarks are expanded on in the Results section.%

The contributions of this work are as follows:
\begin{itemize}
\item A novel neural simulator called EDEN supporting high model generality, computational performance and usability by design.
\item A novel model-analysis/code-generation technique that allows extracting the required calculations from a neural-network model, and casting them into efficient work items that can be run in parallel to simulate the network.
\item A qualitative evaluation of EDEN, demonstrating NEURON-level fidelity, for a diverse set of neural models.
\item A quantitative evaluation of EDEN, demonstrating simulation speeds of real-world neural networks (sourced from literature) up to 2 orders-of-magnitude faster than NEURON, when run on an affordable, 6-core desktop computer.
\end{itemize}

\subsection{Qualitative comparison of neural simulators}

\begin{figure}[t!]{}
    \begin{center}
        \includegraphics[width=0.85\textwidth]{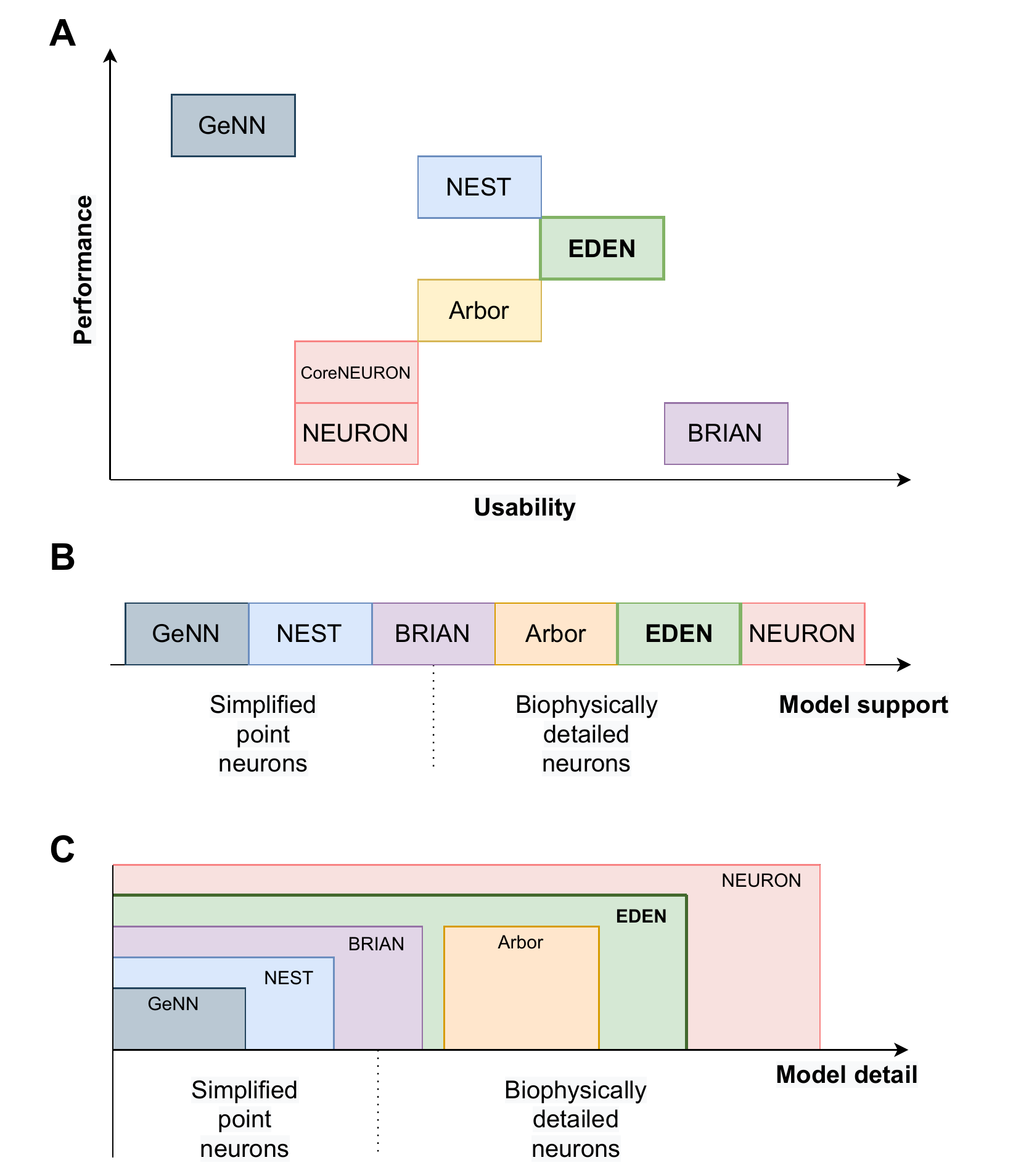}
    \end{center}
    \caption{A relative comparison of the characteristics of EDEN and the established neural simulators. \textbf{(A)}~compares the simulators on the performance and usability axis, \textbf{(B)}~shows the ordering between simulators regarding the level of model detail, and \textbf{(C)}~shows the range of model detail covered by each simulator.}
    \label{fig:pareto_front}
\end{figure}

In Table~\ref{tab:qualitative-comparison}, we present a qualitative comparison between our proposed simulator EDEN, and the most popular, actively developed simulators in the computational-neuroscience field. In line with the scope of this paper, we consider the more general-purpose simulators that can be used in a batch-mode, brain-modelling setting. The table consists of two parts, the top half dealing with coverage of neuron models and features, and the bottom half dealing with aspects of computational performance. Figure~\ref{fig:pareto_front} also summarises a qualitative comparison between the usability, range of supported models and computational performance of the simulators. The characteristics and relative advantages of each simulator are further laid out in the following paragraphs.

\bgroup
\def\arraystretch{1.5}
\begin{table*}[t!]
    \small
    \centering
    \begin{tabular}{|C{3.8cm}|C{1.4cm}|C{1.7cm}|C{1.4cm}|C{1.5cm}|C{1.5cm}|C{1.5cm}|C{1.5cm}|}
        \hline
        \textbf{Features}               & \textbf{EDEN}  & \textbf{(Core) NEURON} & \textbf{Arbor} & \textbf{jLEMS} & \textbf{BRIAN} & \textbf{NEST}  & \textbf{GeNN} \\ \hline
        \multicolumn{8}{|c|}{\emph{Supported Models and Features}}\\ \hline
        \textbf{LIF, AdEx, Izhikevich cells}                    & \yes      & \yes  & \inpart Only LIF  & \yes  & \yes                & \yes                       & \yes                        \\ \hline
        \textbf{Custom artificial cells}                        & \yes      & \yes  & \no             & \yes  & \yes                & \inpart Partially via NestML & \inpart Partially via NineML  \\ \hline
        \textbf{Highly detailed multi-compartmental cells}        & \yes      & \yes  & \yes            & \no   & \inpart Not practical & \no                        & \no                         \\ \hline
        \textbf{Native NeuroML support}                         & \yes      & \no   & \no             & \yes  & \no                 & \no                        & \no                         \\ \hhline{|=*{7}{:=}|}
        \textbf{Support for models \& features compared to EDEN} & Baseline  & \yes  & \no             & \no   & \no                 & \no                        & \no                         \\ \hline
        \multicolumn{8}{|c|}{\emph{Performance}}\\ \hline
        \textbf{Machine-wide parallelism}                & \yes     & \inpart Manual   & \yes           & \no & \inpart Only for simple cases & \yes           & \yes            \\ \hline
        \textbf{Cluster-wide parallelism}                & \yes     & \inpart Manual   & \yes           & \no & \no                    & \yes           & \no             \\ \hline
        \textbf{Cluster-wide auto-parallelisation of detailed networks with graded synapses}          & \yes     & \no            & \no            & \no & \no                    & \no            & \no             \\ \hhline{|=*{7}{:=}|} %
        \textbf{Simulation performance compared to EDEN} & Baseline & \no & \no & \no & \no                    & \yes $\dagger$ & \yes $\dagger$  \\ \hline
    \end{tabular}
    \caption{Qualitative comparison between EDEN and other state-of-the-art neural simulators: NEURON~\citep{neuron}, CoreNEURON~\citep{coreneuron}, JLEMS~\citep{lems}, BRIAN~\citep{brian}, GeNN~\citep{Yavuz2016}, NEST~\citep{Gewaltig2007} and Arbor~\citep{arbor}.
    \\$\dagger$ Only for artificial-cell models that NEST and GeNN support.
    }
    \label{tab:qualitative-comparison}
\end{table*}
\egroup

NEURON~\citep{neuron} is the popular standard simulator for biological and mixed-mode neural networks. It supports the richest set of model features
among neural-simulation packages. A characteristic feature of NEURON is that everything about the model can be changed dynamically while the model is being simulated.
This allows simulation of certain uncommon models, but it negatively impacts the simulator's computational efficiency. CoreNEURON~\citep{coreneuron} is a new simulation core for NEURON that improves computational performance and memory usage at the cost of losing the ability to externally alter the model during simulation. It does not affect setting up the simulator and the model, which are still performed in the same way. Due to the underlying architectural design, the user has to add custom communication code to allow parallel simulation with NEURON, though there is ongoing effort to standardise and automate the needed user code~\citep{netpyne}.

Compared to NEURON, EDEN only supports the NeuroML gamut of models. However, EDEN has much higher computational performance that also automatically scales up with available processor cores and computational nodes. Also, setting up a neural network in NEURON requires the connection logic to be programmed in its own scripting language. This is a cumbersome task and, what is more, NEURON's script interpreter is slow and non-parallel, often resulting in model setup taking more time than the actual simulation.
In contrast, EDEN can load networks from any neural-network generation tool that can export to NeuroML, thus leveraging the capabilities and computational performance of these tools.

Another simulator for biological neural networks is Arbor~\citep{arbor} which aims at high performance as well as model flexibility. Its architecture somewhat resembles the object model used by NEURON, which facilitates porting models, written in NEURON, to Arbor. However, compared to NEURON, it supports a smaller set of mechanisms.
Regarding mixed-mode networks, modelling artificial cells is difficult; only linear integrate-and-fire (LIF) neurons are readily supported, and the user has to modify and rebuild the Arbor codebase for introducing new artificial-cell types. In addition, neuron populations connected by graded synapses cannot be distributed across machines for parallel simulation, which restricts scalability for cutting-edge biological neural models. Compared to Arbor, EDEN supports about the same range of biophysical models; but EDEN also supports \emph{all types} of abstract-neuron models, while Arbor only supports LIF abstract neurons. This limitation prevents Arbor from supporting many mixed-type networks.
There is also a difference in usability: To set up a network model, the network-generation logic must be captured as Arbor-specific programming code. EDEN, instead, avoids simulator-specific programming by using a cross-tool file format.

In the space of artificial-cell-based spiking neural networks (SNNs), there are various specialised simulators in common use.
jLEMS~\citep{lems} is the reference simulator for the LEMS side of NeuroML v2. It supports custom point-neuron dynamics through LEMS, which itself is an extension of the NineML language. It was not designed for high performance and supports only simplified point neurons. BRIAN~\citep{brian} is a simulator originally designed for point neurons, that focuses on usability and user productivity. It supports custom point-neuron dynamics, written in mathematical syntax. Its support for multi-compartmental cells is a work in progress; currently, all compartments must have the same set of equations.
NEST~\citep{Gewaltig2007} and GeNN are general-purpose simulators for networks of point neurons and achieve high performance through a library of optimised codes for specific neuron types. Setting up the network is done through a custom programming language for NEST, and by extending the simulator with custom C++ code for GeNN. For NEST and GeNN, the way to add custom point-neuron types without modifying the C++ code is by writing the neuron's internal dynamics in a simulator-specific language; however, this method is not enough to control all aspects of the model (such as multiple pre-synaptic points on the same neuron, or type of plasticity dynamics respectively). While NEST supports graded synapses, GeNN's architecture is limited to action-potential-based synapses. Compared to abstract-cell simulators, EDEN has an advantage in model generality, since it also supports biophysically detailed multi-compartmental neurons, and hybrid networks of physiological and abstract cells. Although EDEN is not as computationally efficient as the high-end abstract-cell simulators, it readily supports \emph{user-defined dynamics} inside the cells and synapses, whereas said high-performance simulators have to be modified to support new cell and synapse types. EDEN also supports non-aggregable synapses (i.e. not just types that can be aggregated into a single instance as per \citep{Lytton1996} ), and any combination of synapse types being present on any type of cell; which are also not supported by high-performance abstract cell simulators.

An important point to stress is that, the differences in supported model features, combined with the different model description languages, make it difficult to reproduce the exact same neural network (and its output) across all simulators; this is especially so for biologically detailed models. Thus, although there is much previous work on performance-driven neural simulation, our work is one of the first to directly compare performance with NEURON on physiological models that are drawn from existing literature, rather than employing synthetic ones. This further underscores the point that EDEN is a general-purpose tool that can be readily used with existing NeuroML models as well as in new NeuroML projects.

In literature, the authors of CoreNEURON and Arbor have each reported utilising the cores of a whole High-Performance Computing (HPC) node, to achieve up to an order of magnitude of speedup over NEURON. While the models and the machines used in these cases are not the same to allow a strict comparison, our demonstrated speedup of 1$\sim$2 orders of magnitude over NEURON on a 6-core PC, show that EDEN is more than competitive with the state of the art in computational performance. Furthermore, compared to other works on high-performance simulators, EDEN's performance results were measured on a common desktop computer, which makes the results highly relevant for an average neuroscientist's computational resources.

\section{Methods and Materials}

\subsection{EDEN overview}

\begin{figure}[t!]{}
    \begin{center}
        \includegraphics[width=0.4\textwidth]{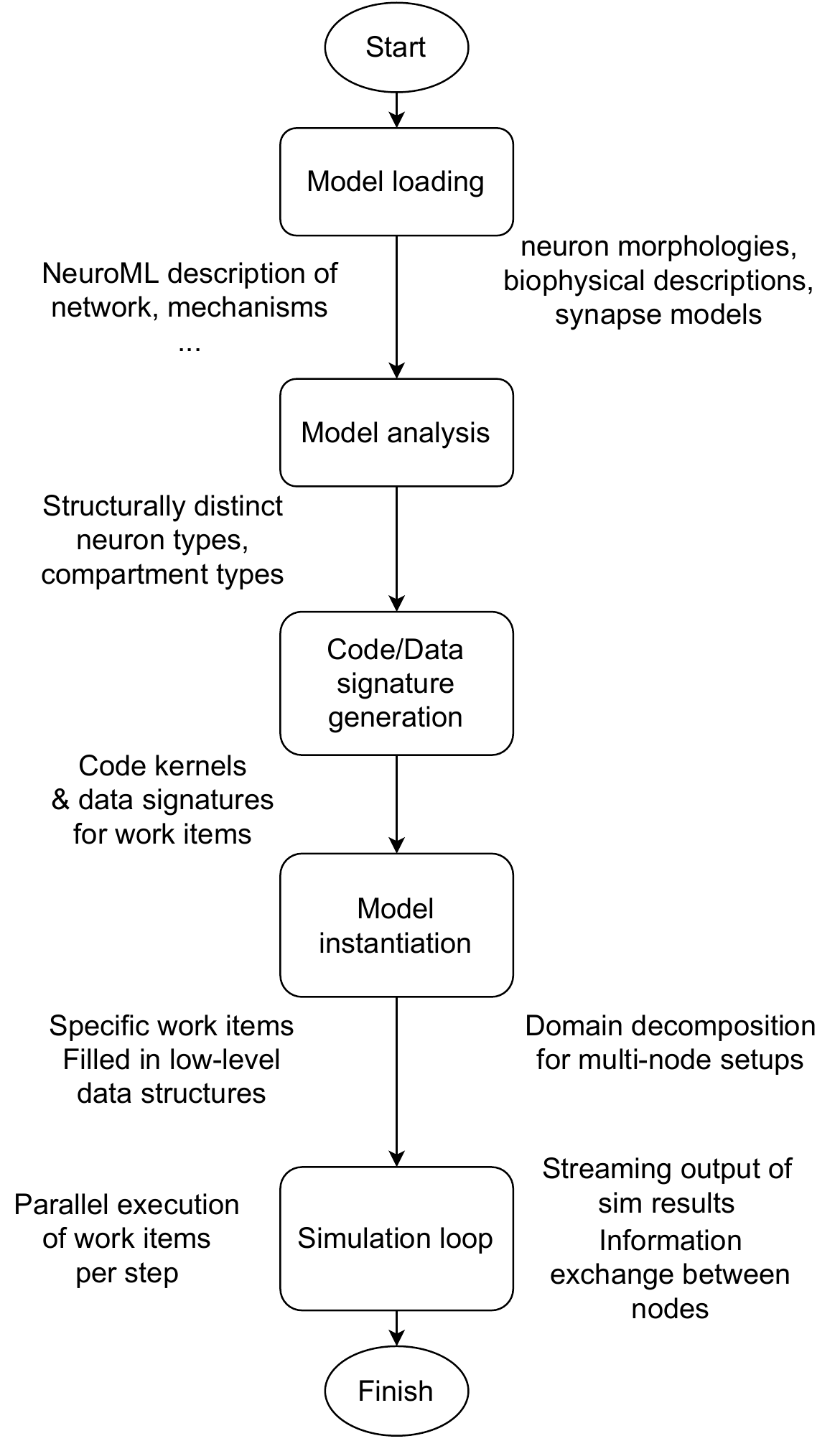}
    \end{center}
    \caption{EDEN's processing pipeline. The whole model is analysed in order to extract the computationally similar parts of neurons, and to generate optimised code and data representations for them, on the fly.}
    \label{fig:eden_flowchart}
\end{figure}

The architecture of EDEN can be visualised as a processing pipeline, which is illustrated in Figure~\ref{fig:eden_flowchart}. The pipeline details and the reasons for enabling EDEN to score high performance, model flexibility and usability are explained in this section. While current neural simulators primarily focus on either computational performance or model generality, EDEN simultaneously achieves both objectives with a novel approach: it generates efficient code kernels that are tailored for the neuron models at hand.

EDEN performs time-driven simulation of 
any sort of neural network that can be
described in NeuroML. To enable simulation of complex, and often heterogeneous, networks with high performance, EDEN first performs model- and workload-analysis steps so as to divide the simulation workload into independent, parallelisable components and, subsequently, determines \emph{efficient code and data representations} for simulating each one of them. Finally, EDEN employs automatic code generation to convert these components to parallel-executable tasks (called work items). Code generation boosts computational efficiency by adapting performance-critical code to the specific model being simulated and to the specific hardware platform being used. Task parallelisation boosts computational efficiency even further by distributing the simulation work across multiple CPU cores in a given computer, and across multiple computers in a high-performance cluster.

\subsection{Usability through native NeuroML support}

Choosing NeuroML as our simulator's input format allows us to focus on our core part of high-performance numerical simulation and, at the same time, leverage the existing NeuroML-compatible, third-party tools for design, visualisation and analysis of neural networks.
Adopting the standard also improves the simulator's usability, as the end user does not need to learn one more simulator-specific modelling language.
In practice, directly supporting the NeuroML standard also allowed us to verify the simulator's results against the standard NEURON simulator, for numerous available models.
As we will see in Section~\ref{sec:results}, the same NeuroML description can be used for both simulators and run automatically. Otherwise, porting all these models separately to both simulators would have taken an impractical amount of effort, making verification and comparison much more difficult to achieve.

\subsection{Performance and flexibility through code generation}

Neural models, especially biophysical ones, are commonly described through a comprehensive, complex hierarchy of mechanisms.
Neural-simulation programmers have to consider this cornucopia of mechanisms and their combinations so as to form neural models. All the while, the formulations behind the mechanisms are constantly evolving, thus, allowing for no single set of mathematical equations to cover most (or even a few of the) neural models.

The resulting complexity -- in both setting up a model and running the simulation algorithm -- has steered general-purpose neural-simulation engines to adopt \emph{object-oriented models} of the neural networks being run. Each type of programming object, then, captures a respective physiological mechanism, and the hierarchy of mechanisms in the model is represented by an equivalent object hierarchy. By adopting NeuroML, EDEN takes the same object-oriented approach at the model input.
Although this approach does help simplify the \emph{programming model} by mitigating the conceptual and programming complexity of working with sophisticated models, it is detrimental to the \emph{execution model}
since it is an inefficient way to run
the simulations on modern computer hardware.
The object-oriented data structure of a model in use has to be traversed hierarchically, every time the equations of the model are evaluated and the model's state is advanced. This, in turn, means that many computations have to be done in a strict order due to the underlying traversal logic, instead of being performed in parallel. Also, the object-oriented model's pointer-based data structures make control flow and data-access patterns unpredictable, slowing down the processing and memory subsystems of the computer respectively.

HPC resources are designed so that the maximum amount of computations can be done independently and simultaneously. Thus, fully utilising them requires streamlined algorithms and \emph{flat data structures}. In many cases, neural-simulation codes have been custom-tailored for the HPC hardware at hand. %
Although such codes improve simulation speed and supported network size by orders of magnitude compared to general-purpose simulators, they make inherent model assumptions that \emph{prevent} them from supporting other models.
The result is that these manually optimised codes, as well as the knowledge behind them, are abandoned after the specific experiment
they were developed for is concluded.
To avoid the pitfalls of these two approaches, EDEN consciously refrains from imposing a specific execution model, so that it can support both model generality and high-performance characteristics. Both of them are simultaneously achieved through a novel approach: efficient code kernels that are tailored for the neuron models at hand are automatically generated, while supporting the whole NeuroML gamut of network models. The specific processing stages that EDEN undergoes to achieve this (see Figure~\ref{fig:eden_flowchart}) are as follows:

\begin{enumerate}[i. ]
    \item Analyse all types of neurons in a given model.
    \item Deduce the parts of the neural network that are structurally similar.
    \item Produce efficient code kernels, each custom-made to simulate a different part of the network.
    \item Iteratively run the code kernels to simulate the network.
\end{enumerate}

This code-generation approach used by EDEN has manifold benefits: First, the simulation can be performed without traversing the model's hierarchy of mechanisms at run time, since the set of required calculations has already been determined at setup time. Second, since the generated code contains only the necessary calculations to simulate a whole compartment or neuron,
the compiler is given much more room for code optimisation compared to code generation for individual mechanisms.
Third, the minimal set of constraints that EDEN's backend places on the code of work items
allows incorporating hand-written code kernels that have been optimised for specific neural models. This is also made possible due to the model-analysis stage, which isolates populations with structurally identical neurons and compartments; when a hard-coded kernel is available for a detected neuron type, it can be employed for the specific cell population, to further boost performance. Thus, EDEN's model-decomposition and code-generation architecture delivers high computational performance for a general class of user-provided neuron models, and it also permits \emph{extensions} in both the direction of model generality and computational performance.

For this first version of EDEN, a \emph{polymorphic kernel generator}\footnote{Polymorphic means that it adapts to the neuron's structure, instead of handling just one type.} that supports the full gamut of NeuroML models was implemented. The specifics of the code kernels are customised for each neuron type; still, the generator's format covers any type of neuron, whether it is a rate-based model, an integrate-and-fire neuron or a complex biological neuron, or whether the interaction is event-based, graded, or mixed. Thus, this implementation provides a baseline of computational efficiency, for all neural models. It can also work in tandem with specialised kernels.
Two ways of extending EDEN with such specialised high-performance codes are described below.

The simplest way to integrate an existing code in EDEN is to directly use it just for the models that the code supports. Programming-wise, the neural network to be run is checked whether it can be run on the new code, and if this is the case, the original new code is generated as a work item, and the simulation data is accordingly allocated and initialised for the model. By running the same code on the same data, extended EDEN should perform as well as the original EDEN code, for the supported family of models.

Alternatively, if the specialised code applies to only a part of the desired network, it can interface with work items from EDEN's general-purpose implementation (or other extensions) for the part of the network that it does not cover. Some modification is then necessary to make the code exchange information (such as synaptic communication) in the same way as the work items it is connected to, but the gains in model generality are immediate.

Following these methods, the usefulness of the optimised code is extended with the least possible effort, simulation can utilise multiple computational techniques at the same time, and the details of each technique do not affect the rest of the EDEN codebase.

\subsection{EDEN concepts}

\subsubsection{Work items}

The fundamental units of work executed per each simulation step in EDEN are called 'work items'. The work items are parts of the model that can be processed in parallel within a simulation step, to advance the state of the simulated model. Within a time-step, each work item is responsible for updating a small part of the entire model. Each work item is associated with a single part of the model data being simulated, and a single code block being run. That code block is responsible for updating the mutable part of its model data over time, but it may also update other parts as well, so that it can send information to other parts of the model. One such case is transmission of spike events to post-synaptic components. Then, data-access collision with the work item that is assigned to the post-synaptic component is avoided by double buffering; the work item receiving the information reads it on the next time-step, while leaving the alternate buffer available for other work items to write to. In the case multiple other work items may write simultaneously, atomic memory accesses are used.

In this first version of EDEN, each work item involves simulating exactly one neuron, but the design allows further variations - for example, to consolidate simple neurons in batches, or to split large neurons in parts  - as long as the calculations for each work item are independent.

\subsubsection{Code and data signatures}

EDEN generates compact code and data representations to run the simulation, by composition of the multiple underlying parts. The details of how this works are explained below.

Each simulated mechanism is defined by its dynamics, the fixed parameters and state variables of the dynamics, and the variables through which it influences other mechanisms, and is influenced by other mechanisms. The external variables influencing the mechanism are called \textit{requirements}, and the values it, in turn, presents for other mechanisms to use, are called \textit{exposures}. Then, to simulate the mechanism, the required actions are to:

\begin{itemize}
    \item evaluate all variables involved in the dynamical equations (called 'assigned' henceforth, in EDEN as well as NEURON parlance). This is the 'evaluation' step of the simulation code.
    \item then, to advance the simulation's state based on the dynamics, and the current values of the assigned variables. This is the 'update' step of the simulation code. %
\end{itemize}

\begin{figure}[t!]{}
    \begin{center}
        \includegraphics[width=1.0\textwidth]{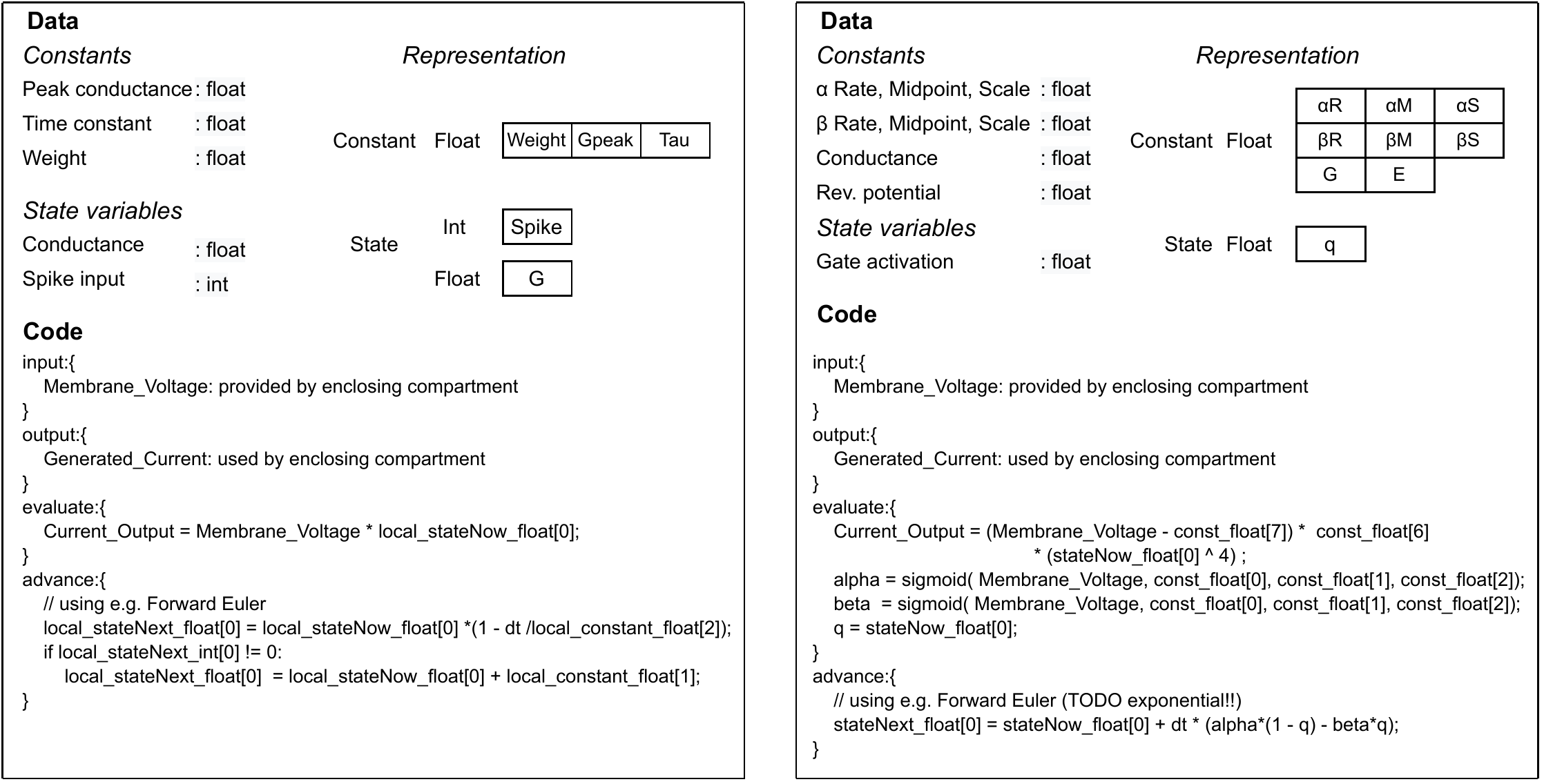}
    \end{center}
    \caption{Code and data signatures for an exponential post-synaptic component \textbf{(A)}, and for a classic Hodgkin-Huxley sodium channel \textbf{(B)}.}
    \label{fig:component_signatures}
\end{figure}

The whole set of code and data to simulate a mechanism is collectively called a \textit{signature} in EDEN parlance. Examples for code-data signatures are shown in Figure~\ref{fig:component_signatures}, for simple cases of a post- synaptic component and an ion channel. Each of them consists of the code to run the 'evaluation' and update' steps (also called \textit{code signature}), and the data representing the mechanism (also called \textit{data signature}). The signature representation is used in EDEN both for simple mechanisms, and composite ones. In fact, the signatures of smaller mechanisms are successively merged to form the signatures of higher-order parts of the neurons, eventually forming signatures for whole compartments or even entire neurons. The code of such signatures is then run in parallel, in order to simulate the whole neural network.

In order to combine the signatures representing two mechanisms, the interfaces(i.e. requirements and exposures) through which the mechanisms interact have to be determined. The hierarchical structure of the provided neural models helps in this, since it delineates the interfaces through which the `parent' mechanism interacts with its `children', and the `siblings' interact with each other.

Code generation starts from the simple, closed-form mechanisms present (for example, Hodgkin-Huxley rate functions or plasticity factors of mechanisms). The hierarchy of mechanisms present in a neuron is traversed, and signatures are incrementally formed for each level of the hierarchy.

The specific steps to merge two signatures are then, in terms of code and data:
\begin{itemize}
    \item The 'evaluation' parts of the code signature have to be placed in a certain order, such that after the variables each mechanism requires are defined and evaluated before the mechanism's evaluation code.
    \item The 'update' parts of the code signature can be appended anywhere after the mechanism's corresponding evaluation code.
    \item The data signatures of the mechanisms are simply concatenated to each other.
\end{itemize}

As signatures  are generated for each higher or lower level mechanism, an auxiliary data structure that has the same hierarchical structure as the original object-oriented model is also formed. This is called the \textit{implementation} of the signature, and it keeps track of how the conversion to signatures was performed, for each mechanism. Relevant information includes the specific decisions made for the generated code (like selection of ODE integrator for the particular mechanism), and the mapping of abstract parameters and state variables (such as the gate variable of an ion channel, the fixed time constant of a synapse, the membrane capacitance of a compartment, etc.) to the specific variables allocated in the data signature. The information is useful for:

\begin{itemize}
    \item referring to parts of the network symbolically (like when recording trajectories of state variables, and when communicating data-dependencies between machines in multi-node setups),
    \item initialising the data structures through the symbolic specifications in use (such as weights of specified synapses),
    \item properly combining signatures, according to implementation decisions (e.g. adjusting the update code to the integrator in use).
\end{itemize}

\subsubsection{Data tables and table-offset referencing}

To achieve high performance during simulation, EDEN uses a simplified data structure for the model.
The model's data are structured in a set of one-dimensional arrays of numbers. These arrays (called \emph{tables} henceforth) are grouped by numerical type (such as integral or floating-point), and mutability (whether their values remain fixed along the simulation, or they evolve through time). This means that each value in the model being simulated is identified by the table it belongs to, its position in the table, and the value's numerical type and mutability.
The value's location can then be encoded into an
integer, from the table's serial number and the offset on the table. The code generated by EDEN can use such references to values at run time, to access data associated with other work items. This relieves EDEN's simulation engine from the need to manage communication between parts of the model with a fixed implementation. Instead, control is given to the generated code on how to manage this communication effectively.

Another benefit of the table-offset referencing scheme is that the references can be redirected to any location in the model's data, if need arises. This is used in particular when a model is run on a computer cluster, where parts of the network are split between computers. In this case, only a fraction of the model is realised on each machine, and the data read by or written to remote parts of the network are redirected to local mirror buffers instead. The change is automatically applied by editing the references in the instantiated data, hence there is no need to change the generated code for the work items.

\subsection{Implementation}

The present implementation of EDEN takes as input NeuroML and supports all neural models in the NeuroML v2 specification.
This implementation, and the code kernels it generates, can be used as a fall-back alternative to further extensions:
the extensions can provide specialised implementations for specific parts of the neural network, while the rest of the network is still covered by the fully general, original implementation.

\subsubsection{Structure of the program}

To begin analysis and simulation of a neural network, its NeuroML representation, along with additional LEMS components describing the custom neuron mechanisms present, is loaded into an object-oriented representation.

The main steps of the process are:

\begin{enumerate}[i. ]
\item Model analysis
\item Work-item generation through code and data signatures
\item Model simulation in the EDEN simulation engine
\end{enumerate}

These steps are further described in the following sections.

\subsubsection{Model analysis and code generation}

The first part in model analysis is to associate the types of synapses in the network with the neuron types they are present in.
This resolves which types of neurons contain which kinds of synaptic components, and where each kind of synapse is located on the neuron.
The same assessment is also made for the input probes connected to each neuron, since probes are also part of the neurons' models.

Then, each neuron type is analysed, to create a signature for each kind of neuron.
First, the structure of the neuron is split into compartments, and the biophysical mechanisms applied over abstract groups of neurite segments are made explicit against the set of compartments.
Thus, for each compartment, we get the entire list of biophysical mechanisms existing on it.
Using these lists, the corresponding code and data signature is formed for each individual compartment.

\begin{figure}[t!]{}
    \begin{center}
    \includegraphics[width=1.0\textwidth]{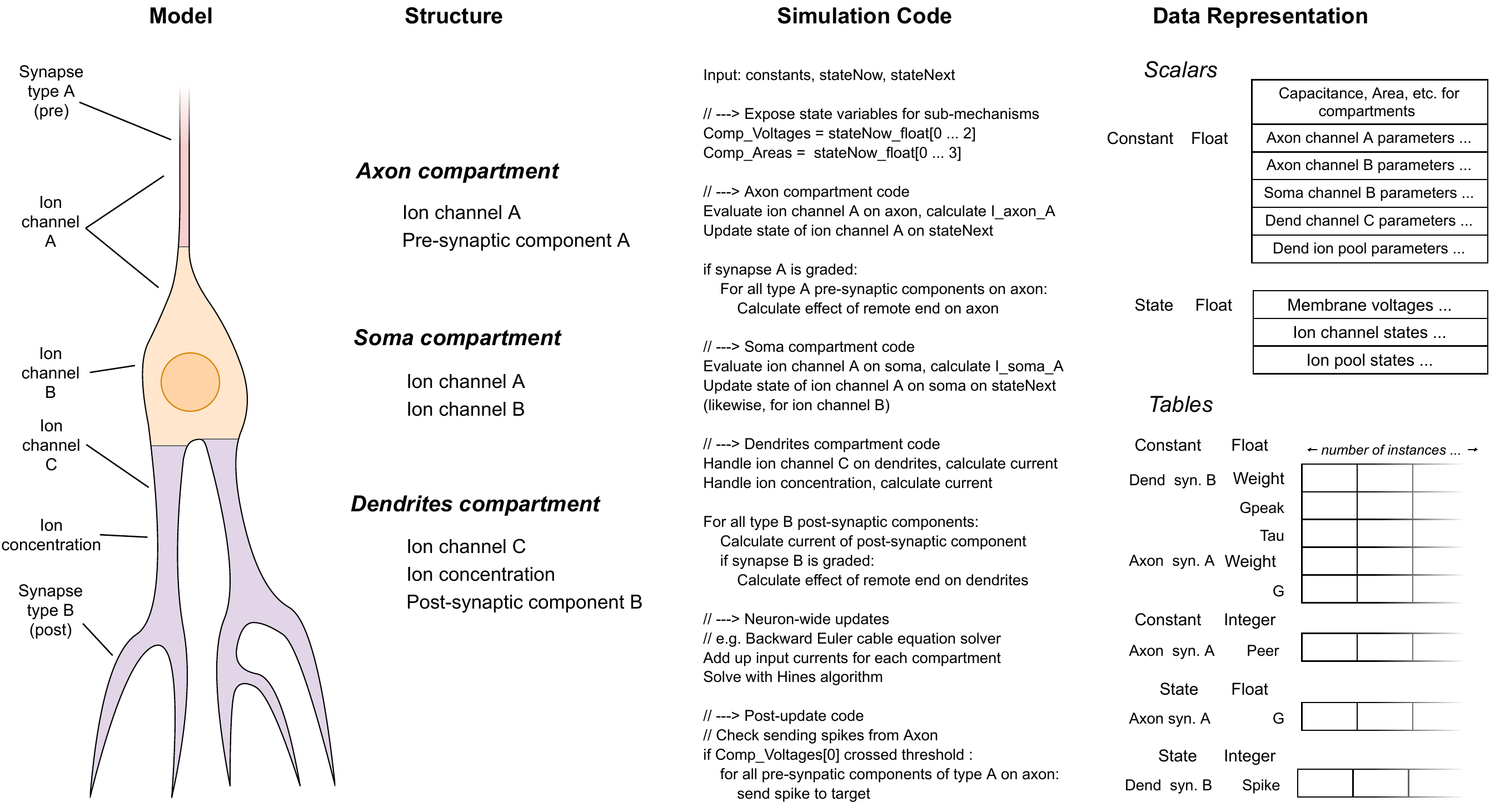}
    \end{center}
    \caption{
    The stages of the per-neuron signature synthesis process, for neurons with few (phenomenological) compartments.
    The neuron shown consists of three different compartments, each containing different physiological mechanisms.
    The simulation code for all mechanisms is laid out in a flat format, along with their associated data.
    Thus, a streamlined and compact code kernel is created for this specific type of neuron.
    }
    \label{fig:neuron_signature_fused}
\end{figure}

If the number of compartments is small, these signatures are concatenated for all compartments present on the neuron, into a neuron-wide signature. This way, a compact code block is generated, with a form similar to how hand-made codes are written for reduced compartmental neuron models.
The process is illustrated in Figure~\ref{fig:neuron_signature_fused}.

\paragraph{Signature de-duplication for identical compartments}

\begin{figure}[t!]{}
    \begin{center}
        \includegraphics[width=1.0\textwidth]{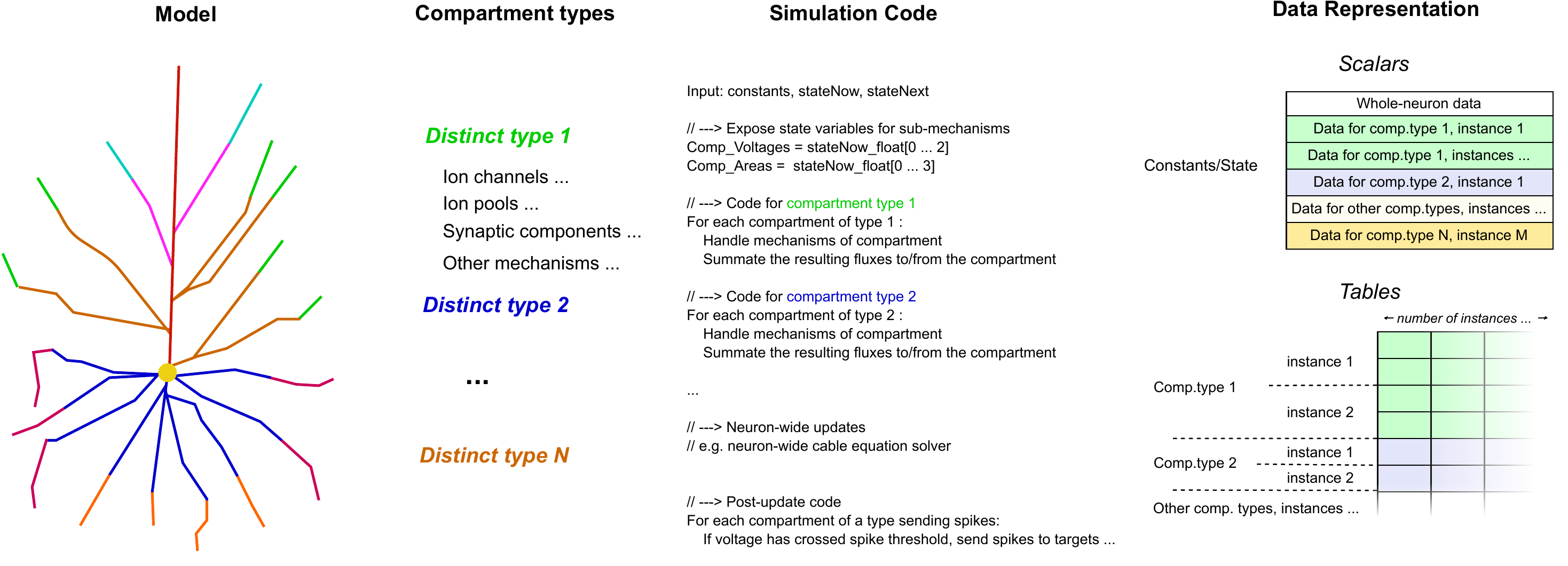}
    \end{center}
    \caption{
        The stages of the per-compartment signature de-duplication process, from the abstract model to the concrete implementation. On the schematic of the detailed neuron model, distinct component types are shown in different colours.
        The components sharing the same type are then grouped together, in terms of simulation code and data representation.
        The specific mechanisms comprising the compartments and the data cells they contain are not shown here, for brevity.
    }
    \label{fig:neuron_signature_deduplication}
\end{figure}

If the amount of compartments is large, it is not practical to generate a flat sequence of code instructions for each individual compartment.
However, in practice, neuron models have less than a few tens of structurally-different sorts of compartments, in the most complicated models.
Thus, a different approach called signature de-duplication is employed, as follows. In this approach, the compartments are grouped for processing, according to their structural similarity (equivalently, similarity of signature representation). The process is illustrated in Figure~\ref{fig:neuron_signature_deduplication}.

Using the per-compartment list of mechanisms, we can immediately deduce which compartments have the exact same structure; which is the case when the set of mechanisms, and thus the code and data signature representation, is the same.
The code signature for the whole neuron now has a set of loops, one for each type of segment. Inside the loop, the code signature to simulate a single compartment is expanded.
Each iteration of the loop performs the work for a different compartment with the same structure. Thus, the data signatures are concatenated together for each group of compartments, and the appropriate offsets are shifted in each iteration of the loop, so that they point to the specific instance of the per-compartment data signature to be used each time.
By generating a specific code block for each sort of compartment, we  eliminate the computational overhead of traversing the sub-components of the mechanisms present on each simulation step, that affects previous general-purpose neural simulators.
Finally, after the code signatures for the work items are determined, they are compiled to machine code, and loaded dynamically on the running process.

\subsubsection{Model instantiation}

After the model is analysed to determine the structure of the work items it is converted to, it is time for the work items and their associated data to be realised in memory.
As mentioned previously, in this version of EDEN, each neuron in the network, along with the synaptic components and input probes attached to it, is assigned to an individual work item. The mapping of parts of the network to work items, is thus fixed.

The data signatures of the work items specify the amount of scalar variables and whole tables each work item uses. Thus, to instantiate each work item, we just have to allocate the same amount of scalars and tables, and keep track of the work item for which these blocks of memory were allocated.
After the variables are allocated for each work item instance, they are filled in, according to the model definition. This is made possible by the implementations of the work items, that keep track of how model-specific references to values map to concrete data values for each work item. Thus, the changes between different instances in the specified model are mapped into changes in the low-level data representation.

The scalar values for each instantiated work item are located in contiguous slices of certain tables, which are reserved for each type of scalar values. Other parts of a work item may not have a fixed size every time. This is, for example, the case for synaptic components of a given type; they may exist in multitudes on a compartment of a neuron, and their number varies across instances of the neuron or compartment type. The data for these variable-sized populations is stored in tables; one set of tables per kind of mechanism on the same compartment. That way, although the sizes of each set of tables may vary eventually, the amount of scalars and individual tables required for a work item remains fixed, for all of its instances.

After allocating the scalars and the tables for the model, what remains is to replace default scalar values with per-instance overrides specified by the model where they exist, and to fill in the allocated tables with their variable-sized contents. The customised scalar values and tables pertaining to the internal models of neurons, are filled in while running through the list of neurons specified in the model. The synaptic connections in the network model are also run through, and the corresponding pre- and post-synaptic components are instantiated on the connected cells. More specifically, on each cell, the tables representing the specified synaptic component are extended by one entry each, with the new entries having the values of the scalar properties of the mechanism. The default values for these properties are provided by the data signature of the mechanism, and customised values (such as weight and delay of the synapse) are filled in using the connection list in the model description.

\subsubsection{Simulation loop}

After model instantiation is done, the code blocks and data structures for the model are set up in system memory and ready to run. Communication throughout the network is internally managed by the code blocks, via a shared-memory model. Double buffering is employed to allow parallel updates of the state variables within a time-step, thus all state-variable tables are duplicated to hold the state of the both the old and new time-step as the latter is being calculated.

All that remains to run the simulation, is to repeat the following steps for each simulation time-step:

\begin{itemize}
    \item set the global 'current time' variable to reflect the new step;
    \item execute the code for each work item in parallel, on the CPU;
    \item output the state variables to be recorded in the network, for the new time-step;
    \item alternate which set of state variable buffers is read from and written to, as per the common double-buffering scheme.
\end{itemize}

Parallel execution of the code kernels is managed by the OpenMP multi-threading library. The `dynamic' load-balancing strategy is followed by default, so whenever a CPU thread finishes executing a work item, it picks the next pending one. The synchronisation overhead of this load-balancing strategy is mitigated by the relatively large computational effort to simulate physiological models of neurons, as will be shown in the Results section.

\subsubsection{Running on multi-node clusters}

Apart from the high-performance properties implemented in EDEN for fast simulation on a single computer,
EDEN also supports MPI-based execution on a compute cluster, so as to further handle the large computational and memory needs of large simulations.
To spread the simulation over multiple co-operating computational nodes, some modifications are made to the process described above. In the following, each co-operating instance of EDEN is called a `node'.

At the model-instantiation stage, the nodes determine which one will be responsible for simulating each part of the neural network.
The neurons in the network are enumerated, and distributed evenly among nodes.
To keep a small and scalable memory footprint, in this version of EDEN, each node is responsible for a contiguous range of the enumerated sequence of neurons.
Then, each node instantiates only the neurons it is responsible to simulate, allocating the corresponding scalar values and tables. The parts that pertain only to individual neurons are also instantiated and filled in. But special care has to be taken when instantiating synapses, since they are the way neurons communicate with each other - and the neurons a node is managing may communicate with other neurons, that are managed by a different node. Thus, the instantiation of synapses is performed in three stages:

\begin{enumerate}
    \item an initial scan of the list of synapses, to determine which information is needed by each node from each node during the simulation;
    \item exchange of requirement lists among nodes, so they all are aware of which pieces of information they must send to other nodes, during the simulation;
    \item establishment of cross-node mirror buffers, and remapping cross-node synapses so that they use these buffers, to access the non-local neurons they involve.
\end{enumerate}

To support these stages, a new auxiliary data structure is created on each node. It is an associative array, mapping the identifiers of peer nodes to the set of information that needs to be provided by that node to run the local part of the simulation (called \emph{send list} from now on). A send list consists of the spike event sources and state variables on specific locations on neurons, that the node needs to be informed about to run its part of the simulation.
The kinds and locations for these state variables and spikes, are stored and transmitted using a symbolic representation, that is based on the original model description. For example, a location on a neuron is represented by the neuron's population and instance identifiers, the neurite's segment identifier, and the distance along that segment from the proximal to the distal part.
Using symbolic representations for send lists allows each node to use the most efficient internal data representation for its part of the model, without requiring peer nodes to be aware of the specific data representation being used on each node. The three stages to set up multi-node coordination are further described in the following:

\paragraph{Synapse-instantiation stage}

First, the list of synaptic connections is scanned, and synapses connecting pairs of neurons are handled by each node according to four different cases:

\begin{enumerate}[i. ]
    \item If a synapse connects two neurons managed on this node, it is instantiated, and the tables are filled in just as described above, for the single-node case.

    \item If neither neuron connected by the synapse is managed by this node, the synaptic connection is skipped.

    \item If the local neuron needs to receive information from the remote neuron (as is the case with post-synaptic neurons and those with bi-directional synapses), then the location on the remote neuron and type of data (e.g. spike event or membrane voltage), is added to the send list for the remote node.
    The local neuron's synaptic mechanism is also instantiated using its data signature, however:
    \begin{itemize}
        \item If the synaptic mechanism is continuously tracking a remote state variable (as is the case with graded synapses), the table-offset reference to that variable is set with a temporary dummy value. This entry is also tracked, to be resolved in the final synapse fix-up stage.
        \item If the mechanism receives a spiking event from a remote source (as is the case with post-synaptic mechanisms), the mechanism receives the spike event in one of its own state variables, instead of tracking a remote variable. (This is the same way event-driven synapses are implemented in the single-node case.) The state variable is used as a flag, so custom event-based dynamics are handled internally.
        Thus, this entry has to be tracked, so that its flag can be set whenever the remote spike source sends a spiking event, at runtime.
    \end{itemize}

    \item If the locally managed neuron does not need to receive information, then is it skipped. The need for this node to send information to other peers  will be resolved in the following send-list exchange stage.
\end{enumerate}

\paragraph{Send-list exchange stage}

At this point, the send lists have been determined, according to the information each node needs from the other nodes.
These send lists are then sent to the nodes the data is needed from; sending nodes do not have to know what they are required to send \textit{a priori}. Therefore, the algorithm described in the following also applies to the more general problem of distributed sparse multigraph transposition~\citep{magalhaes2020efficient}.

In the beginning of this stage, each node sends requests to the nodes it needs data from; each request contains the corresponding send list it has gathered. Then, from each node it sent a request to, it awaits an acknowledgement.
While nodes are exchanging send lists, they also participate asynchronously in a poll of whether they have received acknowledgements for all the requests they sent.
When all nodes have received all acknowledgements, this means all send lists have been exchanged, and the nodes can proceed to the next stage.

By using this scheme, information is transmitted efficiently in large clusters: no information has to be exchanged between nodes that do not communicate with each other. This is a scalability improvement over existing methods,
where the full matrix of connectivity degrees among nodes is gathered on all nodes
\citep{MvdVlag2019}~\citep{magalhaes2020efficient}.

\paragraph{Synapse fix-up stage}

After all data dependencies between nodes are accounted for, each node allocates communication buffers to send and receive spike and state-variable information.
The buffers to receive the required information are allocated as additional tables in the data structures of the model.
They are 'mirror buffers' that allow each node to peek into the remote parts of the network they need to.

The table-offset references that were left unresolved in the synapse-instantiation stage because the required information  was remote, are now updated with references to the mirror buffers for the corresponding remote nodes.
This way, the components of cross-node synapses that -- were the simulation run on a single node -- would directly access the state of adjacent neurons, now access these mirror buffers instead. The mirror buffers are, in turn, updated on every simulation step as described in the next section, maintaining model integrity across the node cluster.

\paragraph{Communication at run-time}

After the additional steps to instantiate the network on a multi-node setup, the nodes also have to communicate continuously during the simulation.
Each node has to have an up-to-date picture of the rest of the network its neurons are attached to, to properly advance its own part of the simulation. Thus, the simulation loop is extended with two additional steps: to send local data to other nodes that need them, and to receive all information from other nodes it needs to proceed with the present time step.

The nodes follow a peer-to-peer communications protocol, which resembles the MUSIC specification \citep{Ekeberg2008}. The data sent from each node to a peer per time-step form a single message, consisting of:

\begin{itemize}
    \item A fixed-size part, containing the values of state variables the receiving node needs to observe.
    \item A variable-size part, containing the spike events that occurred within this communication period.
    The contents are the indices of the events that just fired, out of the full list of events previously declared in the send list.
\end{itemize}

During transmission, each data message is preceded by a small header message containing the size of the arriving message; this is done so that the receiving node can adjust its message buffer accordingly.

After receiving the data message, the fixed-size part is directly copied to the corresponding mirror buffer for state variables, while the firing events in the variable-sized list are broadcast to the table entries that receive them. Broadcasting of firing events is performed using the spike recipient data structure that was created in the synapse instantiation stage.

Inter-node communications are placed in the simulation loop, as follows:
\begin{itemize}
    \item In the beginning of the time step, the information to be sent to other nodes is picked from this node's data structures,
    into a packed message for each receiver. Transmission of these packed messages begins;
    \item Meanwhile, the node starts receiving the messages sent by other nodes to this one. Whenever a message arrives, it is unpacked and the contents are sent to  mirror buffers and spike recipients in the model's data.
    \item When messages from all peers for this node are received, the node can start running the simulation code for all work items, while its own messages are possibly still being sent;
    \item Before proceeding with the next simulation step, the node waits until all messages it started sending have been fully sent; so, then, the storage for these messages can be re-used to send the next batch of messages.
\end{itemize}

\section{Results}\label{sec:results}

During development of the EDEN simulator, we ran functional and performance tests of the simulator, using NeuroML models from the existing literature. The functional tests were used to ensure that the simulator properly supports the various model features specified by NeuroML, and that its numerical techniques
are good enough, with regard to stability and numerical accuracy. The performance tests were done on large neural networks, to evaluate EDEN's computational efficiency and scalability, in various realistic cases.

The NeuroML-based simulations used in the experiments here were sourced from the OSB model repository~\citep{Gleeson2019}, and from NeuroML-DB~\citep{Crook2014}. They were selected to cover a wide range of models in common use (regarding both level of detail and model size), and because their results clearly show various features of neural activity, and how each simulator handles them.
Both simulation accuracy and performance characteristics were compared to the standard NeuroML simulation stack for biophysical models: the NEURON simulator (version 7.7), with the NeuroML to NEURON exporter jNeuroML (version 0.10.0).
Model-porting complications were thus avoided by using the same NeuroML model descriptions.

\subsection{Qualitative evaluation}

To assess functional correctness, i.e. that the various aspects of neural models are simulated properly and accurately, the simulations involved single cells, and small networks with easily discernible characteristics. The specific models in use were gathered from the Open Source Brain code repository; summary details for each model can be found in Table~\ref{table:func-list}. The types of models are further described below.

{
\def\arraystretch{1.5}
\begin{table*}[t!]
    \centering
    \small
    \begin{tabular}{llrr}
        \toprule
        \textbf{Simulation} & \textbf{Type} & \textbf{Compartments} & \textbf{Network} \\
        \midrule
        Izhikevich 2007 & Simplified & 1 & no \\
        FitzHugh-Nagumo & Simplified & 1 & no \\
        Moris-Lecar & Simplified biological & 1 & no \\
        Hindmarsh-Rose & Simplified biological & 1 & no \\

        MainenEtAl simplified & Extended HH & 1 & no \\

        Amacrine & Passive & 2105 & no \\
        Ferrante 2009 & Extended HH & 1091 & no \\
        MouseLight AA0173 & Classic HH & 850 & no \\
        OSB GPUshowcase & Linear I\&F  & 1 & 10 cells \\
        Maex et al. GCL & Extended HH & 1 & 90 cells \\
        \bottomrule
    \end{tabular}
    \caption{Features of the simulations used to test EDEN's functional correctness..}
    \label{table:func-list}
\end{table*}
}

In each case, EDEN's simulation results were checked against the ones produced by NEURON for the same model. NEURON is the most commonly used general-purpose neural simulator, its numerical algorithms have been proven through decades of use, and it is also the one with the most complete NeuroML support to date. The line plots and analogue raster plots in the following figures show the results for each of the simulations run, for the two simulators. Line plots use specific colours for each state variable recorded; they use a darker variant for the trajectories NEURON generates, and overlay a paler variant for the ones EDEN does. When the results match, only the paler colour is visible for the corresponding line.

\begin{figure}[t!]{}
    \begin{center}
        \includegraphics[width=1.0\textwidth]{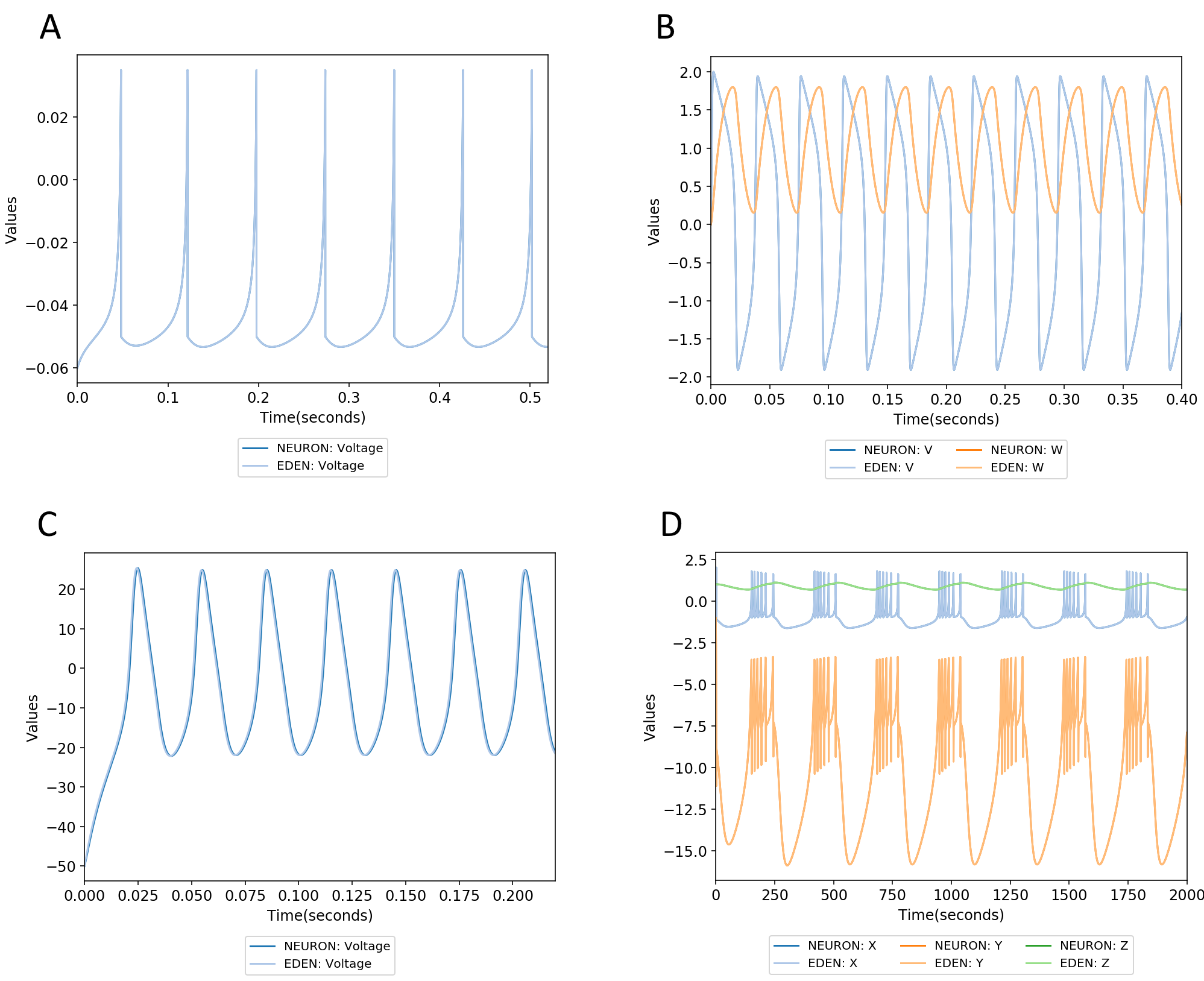}
    \end{center}
    \caption{
        Validation results for various point-neuron types. The trajectories produced by NEURON and EDEN are plotted, with EDEN's trajectories overlaid on NEURON's for each model: Izhikevich 2007 model \textbf{(A)}; FitzHugh-Nagumo model \textbf{(B)};
        Morris-Lecar model \textbf{(C)}; Hindmarsh-Rose model \textbf{(D)}.
    }
    \label{fig:results_func_artificial_cells}
\end{figure}

In Figure~\ref{fig:results_func_artificial_cells}, we show the simulated waveforms for a few common types of artificial cells, namely, the modified Izhikevich model~\citep{izhikevich2007dynamical}, Morris-Lecar~\citep{MORRIS1981}, Hindmarsh-Rose~\citep{HindmarshRose1984} and FitzHugh-Nagumo~\citep{FITZHUGH1961} models. In each model, the neuron receives a DC current input, which induces a periodic spiking pattern. In these plots, the trajectories produced by EDEN are virtually the same as the ones produced by NEURON. Usually artificial cells have slower, less stiff dynamics compared to physiological formulations; so, for that type of neurons, the simulations may not be so sensitive to the choice of numerical integrator.

\begin{figure}[t!]{}
    \begin{center}
        \includegraphics[width=1.0\textwidth]{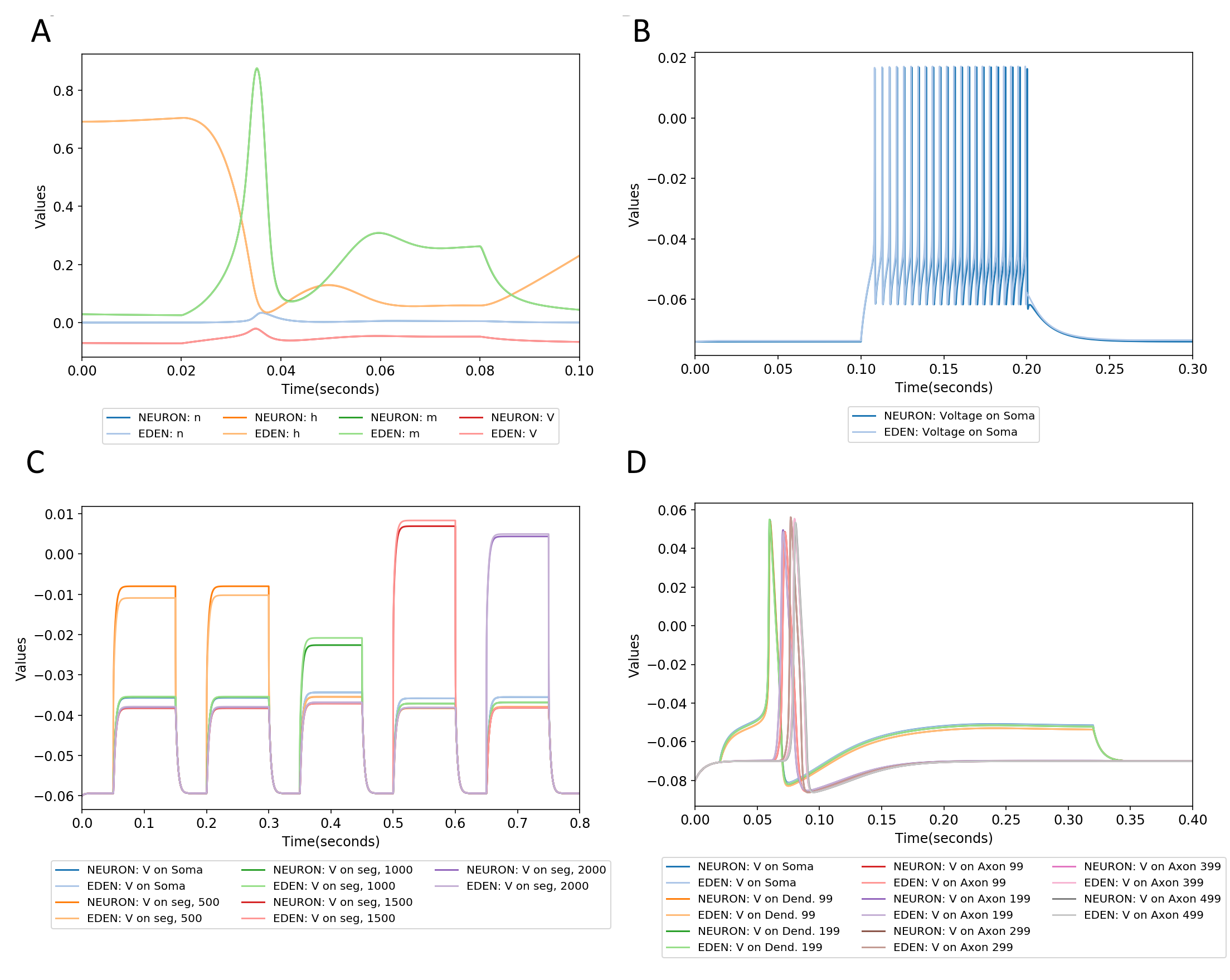}
    \end{center}
    \caption{
        Validation results for various multi-compartmental neuron types. The trajectories produced by NEURON and EDEN are plotted, with EDEN's  overlaid on NEURON's for each model: Mainen  et al. 1995 single-compartment version model \textbf{(A)}; Ferrante 2009 Dentate Gyrus granule cell model \textbf{(B)}; Amacrine passive model \textbf{(C)};
        MouseLight AA0289 model \textbf{(D)}.
    }
    \label{fig:results_func_detailed_cells}
\end{figure}

Figure~\ref{fig:results_func_detailed_cells}A shows simulation results for another kind of point neuron. This is a pyramidal cell model from~\citep{Mainen1995}, but the spatial details of the neuron have been reduced to a single-compartment description. Stimulus sources were not given in the OSB-provided version model, so the neuron fires a transient spike and quiesces.
Figure~\ref{fig:results_func_detailed_cells}B shows simulation results, for a single Dentate Gyrus Granule cell, as described in~\citep{Ferrante2009}. This simulation tracks membrane potential for the soma of the neuron, although the neuron comprises of 1091 anatomical compartments.
We observe that a slight difference in timing occurs between the spike trains generated by NEURON and EDEN. However, this difference is much smaller than a spike period and evolves over a time course of 100 milliseconds. Due to the stiffness inherent in spiking-neuron models, this small delay is enough to allow or inhibit the last spike, the moment when input stimulus stops.

Figure~\ref{fig:results_func_detailed_cells}C shows simulation results, for a passive mouse-retina amacrine cell~\citep{Chen2014}.
The output shows the effect of DC-current probes, placed in different places on the neuron.
This model is based on the reconstructed morphology of the neuron, and physiological data are not available. Thus the NeuroML model includes only passive electrical leaks in the membrane. Here, we can see that there are small steady-state variations in membrane potential, between NEURON and EDEN. Another interesting observation is that these variations happen only on the same compartment that the active DC probe is located at, each time.
Since only membrane leak dynamics are present in the cell, the difference is due to the fact that cell models are discretised differently in the simulators. So, when a DC current flows in the cell from the probe, the compartments have different lengths, so they are converted to different resistor values in the equivalent circuit that is simulated internally.
Thus, the coupling conductance ratio among nodes of the electrical circuit
is different between simulators, which eventually causes the differences in the steady-state potential. As expected, this variation vanishes when we consider the effect of the DC input on the other, more distant compartments. This is because the distances involved are, now, much larger than the length of a single discretisation unit. %

In Figure~\ref{fig:results_func_detailed_cells}D, we see the simulation output for NEURON and EDEN on a reconstructed cell (number AA0173) from the Janelia MouseLight project~\citep{Economo2016}. Since, again, only the passive morphology of the neuron was available, active cell models were produced by adding classical HH ion channels, uniformly distributed across the cell. In this simulation, we observe an action potential caused by a DC current probe on the soma, originating in it and travelling outward along the dendrites. The differences between NEURON's and EDEN's results are very slight.

\begin{figure}[t!]{}
    \begin{center}
        \includegraphics[width=1.0\textwidth]{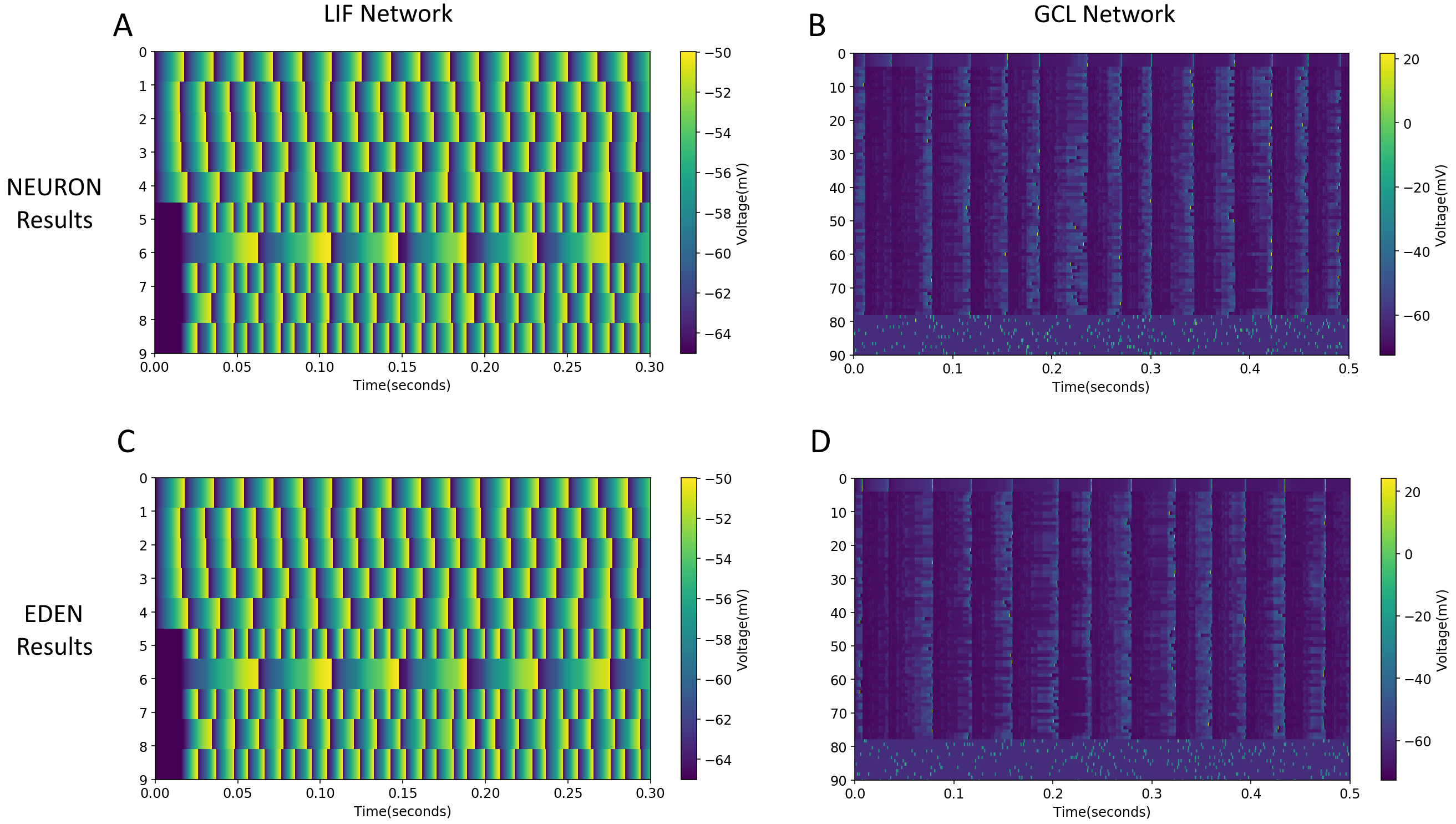}
    \end{center}
    \caption{
        Validation results for the LIF and GCL networks. The trajectories produced by NEURON and EDEN are shown in analogue rasters,
        with NEURON's results on top for each model: LIF model \textbf{(A)} and GCL model \textbf{(B)}; and EDEN's results below, for the same networks \textbf{(C)}, \textbf{(D)}.
    }
    \label{fig:results_func_networks}
\end{figure}

Figure~\ref{fig:results_func_networks} shows the analogue raster plots for two networks of neurons.
The plots show the soma potential of the neurons in the network over time, with the horizontal axis representing time and the vertical axis representing neuron numbers. One network, shown on Figure~\ref{fig:results_func_networks} (A and C) consists of linear integrate-and-fire neurons, connected with double-exponential conductance synapses. It is a test network that is provided by OSB as a target for simulators on accelerator platforms.
Since there is no random-variable component in this simulation, its results are deterministic. The results of the two simulators are virtually identical, just like in the previously shown simulations of isolated artificial cells.

The other network shown on Figures~\ref{fig:results_func_networks} (B and D) contains three interacting populations of physiologically-modelled point neurons, as per the cerebellar granule-cell layer model of Maex et al.~\citep{Maex1998}. The inputs provided to this cell are stochastic (double-exponential firing synapses, triggered by independent Poisson spike sources).

We observe synchronised firing in the group of Golgi cells (neurons 1 to 4), a more variable, roughly synchronous firing pattern in the group of granule cells (neurons 5 to 78)  and random individual firing for the group of mossy fiber cells (neurons 79 to 90). The synchronised firing fronts for the entire network are not periodic.
In fact, the synchronous and individual cell firing varies among runs with different randomisation seeds, on either NEURON or EDEN. Hence, results from any two runs of the model can only be compared through indirect metrics, even then the same simulator is used. However, we observe that the model exhibits similar activity patterns on both NEURON and EDEN.
(Note that the shown raster plots have a fine temporal resolution; the seemingly flat-coloured voltage 'boxes' that are visible are due to sub-threshold excitatory post-synaptic potentials, and their width in time represents the duration of the potentials.)
In conclusion, the above qualitative tests provide an insight on how close EDEN's simulation results are to the ones produced by NEURON, for various cases of neural models. In the case of point neurons, the results produced by both simulators are virtually the same. In the case of multi-compartment cells, the differences in simulation results are minor, and caused by the different discretisation strategies and numerical integration techniques the simulators employ.

\subsection{Performance evaluation}

\subsubsection{Overview}

Beside flexibility in supported models, another distinguishing characteristic of neural simulators is speed.
To evaluate the simulation speed EDEN offers we ran simulations of neural networks available in literature, on a recent cost-effective desktop computer. We chose to run published neural networks over synthetic benchmarks, because:

\begin{itemize}
    \item they have been used in practice, so they are concrete examples of what end users need; and
    \item existing models are usually the base for newer models, so the insights about the former do remain relevant.
\end{itemize}

Since the original neural networks were developed with the computing limitations of earlier years, these days they run comfortably in a desktop computer, using a minor fraction of system memory and within just a few minutes per run. (Unfortunately, new network models that do push the limits of present hardware, are still only available in heavily custom setups, that cannot be easily ported to another data format, simulator, or HPC cluster.)
To evaluate simulation performance for longer simulation run times, and more challenging neural network sizes, we also used enlarged versions of the original neural networks. This was possible, because the original networks were themselves procedurally generated, with parametric distributions of networks and synapses.

The neural networks that were run for performance evaluation are listed in Table~\ref{tab:performance-details}, along with quantitative metrics for each case.
Beside these quantitative metrics, there also are substantial qualitative differences between the models. These differences determine both the neural functions of each network, as well as the required computational effort to simulate each one.

{
    \def\arraystretch{1.5}
    \begin{table}[t!]
        \centering
        \small
        \begin{tabular}{lcrrrr}
            \toprule
            \textbf{Simulation}   & \textbf{Simulated time} & \textbf{Steps}   & \textbf{Neurons} & \textbf{Total Compartments} & \textbf{Total Synapses}  \\
            \midrule
            GCL       & \multirow{2}{*}{1 sec}   & \multirow{2}{*}{40000} &   728 &   6824 &    5475 \\ %
            GCL x10   &                          &                        &  7280 &  68240 &   54933 \\ \cmidrule{1-6} %
            M1 5\%    & \multirow{3}{*}{1 sec}   & \multirow{3}{*}{20000} &   527 &    527 &   15469 \\ %
            M1 10\%   &                          &                        &  1065 &   1065 &   61538 \\ %
            M1 100\%  &                          &                        & 10734 &  10734 & 5032223 \\ \cmidrule{1-6}%
            CGoC      & \multirow{2}{*}{0.1 sec} & \multirow{2}{*}{40000} &    45 &  28170 &    1348 \\ %
            CGoC x10  &                          &                        &   450 & 281700 &   14930 \\
            \bottomrule
        \end{tabular}
        \caption{The simulated networks used for performance benchmarking.}
        \label{tab:performance-details}
    \end{table}
}

\subsubsection{Simulated networks}

The neural network models used were the following, sourced from NeuroML-DB:

\begin{enumerate}
    \item A multi-compartmental extension of the~\citep{Maex1998} Cerebellar Granule Cell Layer (\textbf{GCL}) model (NeuroML-DB ID: NMLNT000001).

    \item An Izhikevich cell-based, multiscale model of the mouse primary motor cortex (\textbf{M1})~\citep{Dura-Bernal2017}  (NeuroML-DB ID: NMLNT001656).

    \item A model of the Golgi cell network in the input layer of the cerebellar cortex(\textbf{CGoC}), electrically coupled with gap junction~\citep{VERVAEKE2010435} (NeuroML-DB ID: NMLNT000070).
\end{enumerate}

\paragraph{The GCL network}

The GCL network is based on the \citep{Maex1998} model for the cerebellar granule cell layer, which includes granule cells, Golgi cells, and mossy-fiber cells. The original model was extended to have multi-compartmental cells; in particular, the axons and parallel fibers of the granule cells are spatially detailed with 11 compartments per cell, and Golgi cells follow the ball-and-stick model, with 4 compartments per cell. The mossy-fiber cells are stimulated by Poisson randomly firing synapses and stimulate the granule cell population through AMPA and blocking NMDA synapses. The granule cells excite the Golgi cells through AMPA synapses, and the Golgi cells inhibit the granule-cell population through GABA\textsubscript{A} synapses.
We enlarged the original GCL network, by multiplying the population size by a factor of 10, and keeping the same  per-neuron synapse density for the various projections. Thus, the total amount of synapses was also 10 times the original.

\paragraph{The M1 network}

The M1 network is an Izhikevich cell-based model of the mouse primary motor cortex, with various groups of cells intertwined across cortical depth. There are 13 groups of cells and 4 different sets of dynamics parameters among the cells. Each cell is stimulated by an external randomly firing synapse stimulus, and cells interact with each other through excitatory AMPA and MNDA synapses, and inhibitory GABA synapses. All synapses follow the stateless, double-exponential conductance model.
This model is rather recent, so in its full size, it is computationally challenging enough to simulate, without enlarging it.

To better investigate performance characteristics, and evaluate performance at a model scale similar to the original GCL and CGoC networks, we generated two smaller versions of the M1 network, at 'scale' values of 10\% and 5\%. Note that the model uses fixed probability connectivity
for the various projections between populations, thus the amount of synapses grows quadratically with the population size.

\paragraph{The CGoC network}
\label{par:cgoc-explain}
The CGoC network models a small part (0.1 mm\textsuperscript{3}) of the Golgi cell network, in the input layer of the mouse's cerebellar cortex. It was used in~\citep{VERVAEKE2010435} to investigate the network behaviour of Golgi cells, using experimental data. In this network, the neurons communicate with each other solely through gap junctions. Each cell also has 100 excitatory inputs in the form of randomly firing synapses, randomly distributed among apical dendrites.

Gap junctions have rarely been introduced in large network models in the past. This is not because they are absent from tissue, nor because their effects are negligible, but primarily because of their intense computational requirements. The continuous-time interaction between neurons that gap junctions effect requires a large amount of state data to be transferred to simulate each neuron in every step, while spike-based synapses need to only transfer the firing events between neurons, whenever they occur (rarely, compared to the amount of simulation steps).
As with the GCL network, we also enlarged this network, by making neuron count, synapse count and network volume 10 times the original.

\subsubsection{Benchmark results}

The three neural networks described above -- with network sizes, simulation time and time-steps as shown on Table~\ref{tab:performance-details} -- were run on a recent desktop PC, and simulation run time was measured in each case. EDEN's run time was measured both when using all CPU cores of the PC, and when running on a single CPU thread. These two cases reflect different cases of the simulation workload: the first one is when an individual simulation has to be run, and the second one is when a large batch of simulations has to be run, as a group.
The NEURON simulator was chosen as a baseline to compare simulation performance, because it is the predominant simulator for biophysically detailed, multi-compartmental neuron models.

For all performance benchmarks, the machine used was a desktop PC, with a  with a Intel i7-8700 3.2Ghz CPU, and 32GB of 2133 MT/S DDR4 RAM. The CPU has 6 physical cores, and can run up to 12 (hyper)threads simultaneously. The particular CPU was selected, to reflect the typical processor available on a researcher's desk - rather than what is available on a supercomputer setting, which requires substantial technical effort to use, and is often not available for day-to-day experimentation. The OS used was Ubuntu Linux 18.04. NEURON, EDEN and the code generated by both at runtime were all compiled using the GNU C compiler, version 7.4.

{
    \def\arraystretch{1.5}
    \begin{table*}[t!]
        \centering
        \small
        \begin{tabular}{lrrrrr}
            \toprule
            \multirow{1}{*}{ \textbf{Experiment} } & \multirow{1}{*}{ \textbf{Neuron run time} } & \multicolumn{2}{c}{ \textbf{EDEN Run time} }
            & \multicolumn{2}{c}{ \textbf{EDEN Speedup} } \\
            & & 1 thread & Full node & 1 thread & Full node \\
            \midrule
            GCL       &    83.72 &  16.95 &   3.09 & $\times$  4.94 & $\times$  27.08 \\ %
            GCL x10   &  2049.89 & 300.05 &  85.04 & $\times$  6.83 & $\times$  24.11 \\ %
            M1 5\%    &    13.34 &   7.32 &   1.40 & $\times$  1.82 & $\times$   9.52 \\ %
            M1 10\%   &    52.45 &  22.79 &   4.20 & $\times$  2.28 & $\times$  12.48 \\ %
            M1 100\%  &  5423.53 & 889.17 & 364.88 & $\times$  6.10 & $\times$  14.86 \\ %
            CGoC      &   514.96 &  22.72 &   3.51 & $\times$ 22.67 & $\times$ 146.80 \\ %
            CGoC x10  & 12849.54 & 459.26 & 159.13 & $\times$ 27.98 & $\times$  80.75 \\
            \bottomrule
        \end{tabular}
        \caption{Measured run time for benchmarks for NEURON, EDEN on 1 thread, and EDEN using all CPU threads, and respective speedup ratios.}
        \label{tab:performance-results}
    \end{table*}
}

\begin{figure}[t!]{}
    \centering
    \includegraphics[width=1.0\textwidth]{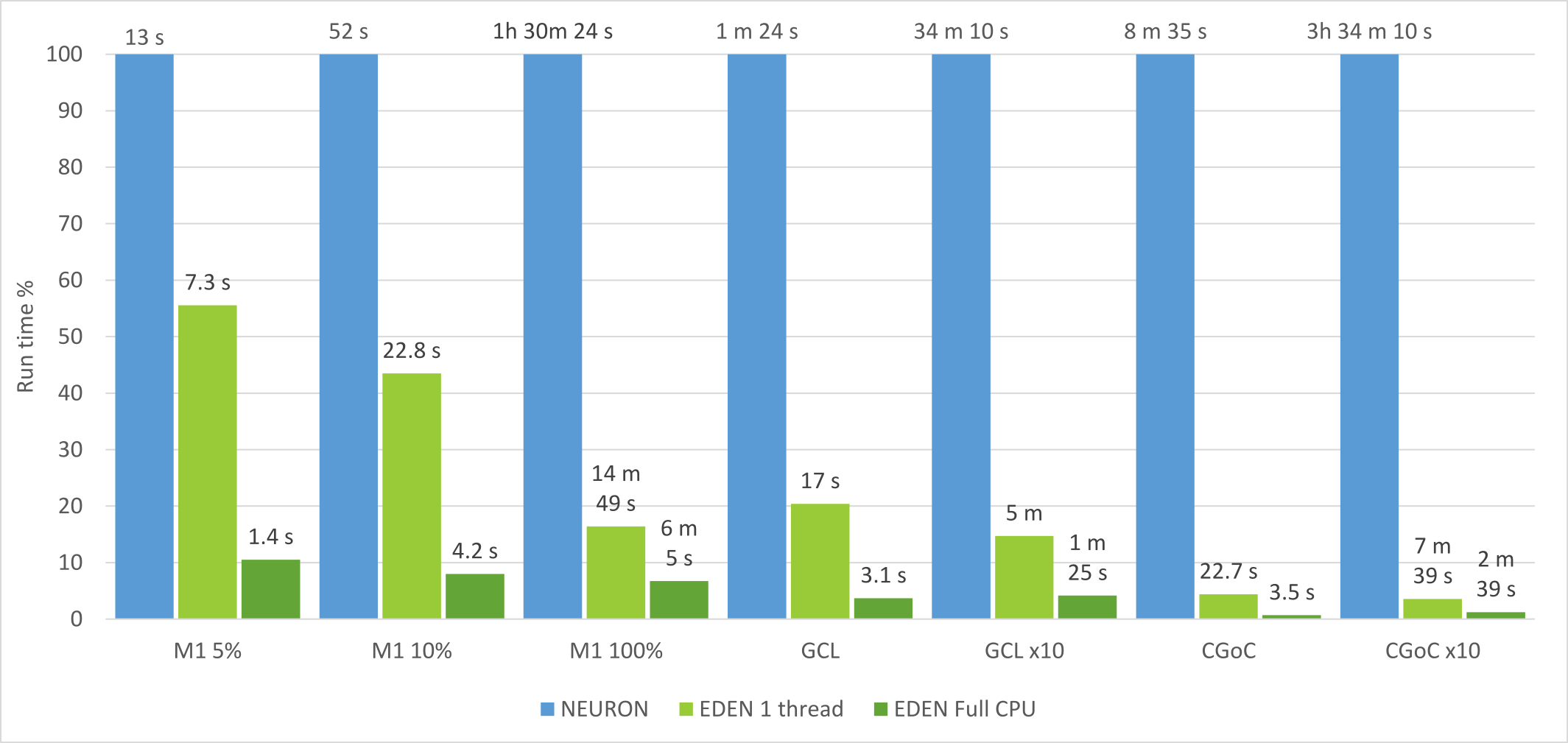}
    \caption{Run time for each neural network considered, for NEURON vs. EDEN on 1 CPU thread vs. EDEN on all CPU threads. For each case of neural network, the bar height in the chart is normalised against NEURON's run time for that case.}
    \label{fig:performance_time_general}
\end{figure}

The results for the performance benchmarks are shown in Table~\ref{tab:performance-results}. For each simulation in Table~\ref{tab:performance-details}, the time to run it is shown when running NEURON, EDEN on one CPU thread, and EDEN on the whole CPU. The corresponding speedup ratios for EDEN on a single thread and for EDEN on all threads over NEURON, are also shown on the table. Figure~\ref{fig:performance_time_general} visualises the relative time to run each simulation with EDEN, using one CPU thread or the whole CPU, against the time to run the same simulation with NEURON.

We observe that EDEN vastly outperforms NEURON while running on a single CPU thread, and even more so when the network is simulated across all threads of the CPU. %
This is because EDEN was designed from the start to achieve high computational performance, especially when running complex, biophysically detailed neurons.
In the following, we will comment on the performance characteristics demonstrated when running each specific neural network, and reiterate the network's properties that affect computational performance.

The GCL network comprises biophysically detailed cells, with a small number of compartments per cell. In this case, EDEN generates fully simplified code kernels for each neuron type; the code to simulate each individual compartment is laid out as an explicit, flat sequence of arithmetic operations.

When running the original GCL network, EDEN runs at 4.9 times the speed of NEURON, using one CPU thread. This level of speedup over NEURON applies when running a batch of many small simulations; in which case, each simulation is run on a single CPU thread for best results. By utilising all 6 cores of the CPU, simulation speed further improves more than five times over, for a total of 27.1 times the speed of NEURON. This shows that when a single simulation has to be run at maximum speed, EDEN can automatically, efficiently parallelise the computational work across multiple processor cores to run faster.

For the enlarged version of the network, single-thread speedup using EDEN increases further, to 6.8 times the simulation speed of NEURON. Speed improves even more by using all threads, but the total improvement in speed versus running NEURON is not as great as when running the smaller, original-size model ($\times$24.1 total, compared to $\times$27.1 previously). A possible reason is that the processor's data transfer speed decreases with model size, and limits computational throughput.

The M1 network comprises Izhikevich-type artificial cells, with dense synaptic connectivity between the neurons. In this case, each neuron's internal model is one Izhikevich-cell mechanism; EDEN generates a simplified code kernel, that simulates the neuron's internal dynamics and synaptic interaction.

When running the 5\% version of the network, EDEN on a single CPU thread runs at 1.8 times the speed of NEURON, and using all cores it runs at 9.5 times the speed of NEURON. Running the larger 10\% network, these performance ratios increase to 2.3 times over and 12.5 times over respectively. Finally, when running the full-sized version of the network, EDEN on one CPU thread runs at 6.1x the speed of NEURON, and using all cores it runs at 14.9 times the speed of NEURON.

For the reduced-size versions of the network, EDEN still runs faster than NEURON, but not by as much as  when running the full-sized version.
This is because the amount of computations and data involving these simplified neurons is smaller, which magnifies the slow-down effects of parallelisation overhead and CPU to memory transfer rate.
For the full-sized network, EDEN's relative performance improves steadily.
Although the performance boost is somewhat reduced compared to when running biophysically-detailed models, it still is substantial for all network sizes. %

Networks solely consisting of point neurons can already be run with high computational performance, on specialised simulators like NEST.
However, there is the important %
class of hybrid SNNs~\citep{LyttonHines2004}, that mixes physiologically-detailed and artificial cells according to the focus of each model.
Such networks have to be run with general-purpose neural simulators, that support both types of neuron in tandem. By demonstrating a consistent high speedup factor even for artificial-cell networks that are not its main target, EDEN shows that it can run hybrid neural networks at a greatly increased speed, without running into performance problems.
For pure artificial-cell networks, EDEN is still relevant for modifications that break out of the commonly supported models, or take a lot of effort to set up on high-performance artificial-cell simulators (e.g. require modifying the simulator's source code to extend model support).

The CGoC network is made up of Golgi cells, which are modelled with hundreds of physiological compartments. Since these cells have too many compartments to apply a flat-code representation per cell type, as was done for the GCL network, EDEN works differently in this case.
For each type of cell, the compartments comprising it are grouped according to the set of physiological mechanisms that they contain. This way, one code kernel is generated to simulate each different type of compartment. Then, all compartments of the same type are simulated as a group, using a loop over the same code. After computing the internal dynamics for each compartment, the interactions between the compartments, such as the cable equations, are also computed to complete the time-step.

We notice that when running either the original or the 10x-enlarged version of the CGoC network, EDEN exhibits a spectacular increase in simulation speed compared to NEURON. When running the original-sized network, the relative simulation speed over NEURON is 22.7 times using one thread, and 146.8 times using all threads. In wall-clock terms, this means that a simulation that used to take eight and a half minutes to run with NEURON, takes 22.7 seconds with EDEN in batch mode, and 3.5 seconds with EDEN in single-simulation mode. When running the 10x-enlarged version of the network, the relative simulation speed using NEURON is 28.0 times when using one thread, and 80.7 times when using all threads. In this case, wall-clock run times are three and a half hours to run with NEURON, versus 7 minutes 39 seconds with EDEN in batch mode and 2 minutes 39 seconds with EDEN in single-simulation mode.

\section{Discussion}

\subsection{Current neural-simulator challenges}

Through the process of developing EDEN and our involvement with the existing neural-modelling literature, tools and practices,
we realised the urgent need for \emph{standards} in brain modelling and \emph{reproducibility} between simulators.
From the perspective of a computer engineer, there is an enormous learning curve in designing simulators for biophysically-detailed neural networks.
The technical know-how on handling the differential equations of neural physiology is scattered across past publications and program source code, and even then rarely mentioned by name.
Having a standard would help as a compendium of all the mathematical concerns that affect programming, and allow neuroscientists and engineers to cooperate without continuous friction.

As mentioned in the Introduction section~\ref{sec:intro}, when working with highly detailed neural networks, to change the simulator in use would take an impractical amount of effort.
This is one of the reasons why there are so few inter-simulator comparisons for the same model in literature,
and why they usually are about porting a custom simulation code to, or from, a general-purpose simulator.  
A standardised, compatible description for models would obviate this major obstacle and enable cross-simulator evaluation.

Another important aspect of upcoming neuroscience projects is \emph{multiscale modelling}, that is, studying a neural structure through multiple levels of modelling detail.
Since this often involves many different simulators of different model types, it is only practical through extensive standards that capture not only the different models but also the results of the simulation at each level.
This is necessary in order to  reconcile and investigate the different scales of modelling without fully custom, one-off code.

Besides standards, we also advocate for a more rigorous \emph{integration} of the various simulators with neuroscientific but also general (e.g., Python/Jupyter) workflows, which will speed up experimental setup as well as seamless transfer of simulation results across different platforms.
This may sound obvious but is in fact a crucial element for real-world quick adoption and utilisation of this ensemble of platforms.
NEURON, BRIAN, GeNN, and Arbor have already caught on to this need; that is why they all natively support a Python interface, alternative to their own custom languages. (BRIAN itself is Python-native.)
EDEN already offers such integration, through the existing NeuroML tooling infrastructure.

Regarding the NeuroML community, it is important to stress the usefulness of providing simulation files along with the published model descriptions.
This is important not only to fully record the published experiments, but also to be able to reproduce the experiments, and cross-validate the results on multiple simulators.
To illustrate, we tried to evaluate EDEN on as many NeuroML networks as possible but were only able to find five individual, non-trivial network models in the entire NeuroML-DB --
and important simulation parameters such as duration, time-step size, and recording probes were only available in the original code repositories outside NeuroML-DB.

Finally, from an HPC perspective, the large-network simulation needs of modern researchers call for the use of computer clusters.
However in the existing simulators, support for clusters is either partial, or requires advanced programming from the end user to work.
Automatic, complete support for clusters must therefore be a development priority, which the simulator authors are best suited to address.
EDEN offers such built-in automation and will continue improving on its performance.

\subsection{The EDEN potential and next steps}

The evaluation presented makes it abundantly clear that EDEN delivers on its triple mission towards high performance, high model generality and high usability. This first version of EDEN was focused on ensuring that all kinds of NeuroML models are supported, rather than optimising the performance of a limited subset.
Thus, the performance results seen in this work form a minimum guaranteed baseline of performance, on top of which future extensions can boost performance even further.

Even so, we showed that this performance baseline provides, for real-world neural networks drawn from NeuroML-DB, a speedup ratio over NEURON of 2$\sim$28$\times$ per CPU thread and 9$\sim$146$\times$ in total, on an ordinary desktop PC. We also demonstrated that no technical expertise is required for deploying and parallelising the simulations of small and large networks alike, which presents a great incentive for the quick adoption of EDEN by the neuroscientific community.

All its achievements notwithstanding, EDEN is far from a concluded simulator. Our future plans involve work in various directions. To enumerate a few crucial ones:

\begin{enumerate}[i. ]
    \item Validate further the EDEN architecture through integrating existing, best-in-class code kernels from the community for special cases. Characterise performance etc. on various types of neural networks so as to determine further performance margins.

    \item Boost the general-purpose EDEN backend by porting it to accelerator hardware, e.g. on GPUs and graph processors. Employ graph theory methods for problem mapping in order to deploy EDEN on heterogeneous (e.g., CPU-GPU) platforms and optimise communication overhead.

    \item Add further extensions to EDEN for high-end HPC application, such as support for the SONATA data format and for simulation checkpointing.

    \item Research and refine innovative numerical integrators, to improve computational parallelism and maintain numerical accuracy on  challenges like cable equations and kinetic schemes.

    \item Evaluate and propose extensions to EDEN and NeuroML that enable direct interfacing with arbitrary data sources such as video stimuli, simulated environments to allow training experiments, and dynamic clamps for hybrid experimentation.
\end{enumerate}

\section{Conclusion}

The large scale, fast pace, and wide diversity of \insilico neuroscience requires simulation platforms that offer high computational performance alloyed with reproducibility, low complexity in model description and a wide range of supported mechanisms.
To those ends, we presented EDEN, a novel neural simulator that natively supports the entire NeuroML v2 standard, manages the simulation's technical details as well as multi-node and multi-core cluster resources automatically, and offers computational performance without precedent in the scope of general-purpose neural simulators.

\section*{Conflict of Interest Statement}

The authors declare that the research was conducted in the absence of any commercial or financial relationships that could be construed as a potential conflict of interest.

\section*{Author Contributions}

S.P. designed and developed the EDEN simulator, and conducted the experiments.
H.S. was the technical manager of the simulator project, and designed the experiments.
M.N. provided guidelines for usability, and for numerical issues of neural simulation.
D.S. and C.S. conceived and supervised the simulator project.
The manuscript was written and edited jointly by all authors.

\section*{Acknowledgements}
This research is supported by the European Commission Horizon2020 Framework Programme Projects EXA2PRO (Grant Agreement No. 801015) and EuroEXA (Grant Agreement No. 754337).

\section*{Data Availability Statement}

The source code of the EDEN simulator is available on GitLab: \url{https://gitlab.com/neurocomputing-lab/Inferior_OliveEMC/eden} .

\bibliographystyle{frontiersinSCNS_ENG_HUMS} %
\bibliography{test}

\end{document}